\documentclass[aps,superscriptaddress,preprint,showpacs,pra,twoside,amssymb,amsmath,latexsym]{revtex4}
\newcommand{\bF}{{\bf F}}
\newcommand{\bZ}{{\bf Z}}
\newcommand{\bX}{{\bf X}}
\newcommand{\ba}{{\bf a}}
\newcommand{\bb}{{\bf b}}
\newcommand{\bc}{{\bf c}}
\def\complex{\mathbb{C}}
\def\Tr{\mathop{\rm Tr}\nolimits}

\def\argmin{\mathop{\rm argmin}}
\newcommand{\rE}{{\rm E}}
\newcommand{\pos}{{\rm pos}}
\newcommand{\cH}{{\cal H}}
\newtheorem{lem}{Lemma}
\newtheorem{thm}{Theorem}
\def\mix{{\rm mix}}

\def\choose#1#2{\genfrac{(}{)}{0pt}{}{#1}{#2}}
\begin{document}
\title{Practical Evaluation of Security for Quantum Key Distribution}
\author{Masahito Hayashi}
\email{masahito@qci.jst.go.jp}
\address{EARTO-SORST Quantum Computation and Information Project, JST\\
5-28-3, Hongo, Bunkyo-ku, Tokyo, 113-0033, Japan\\
Superrobust Computation Project,
Information Science and Technology Strategic Core (21st Century COE by 
MEXT)\\
Graduate School of Information Science and Technology,
The University of Tokyo\\
7-3-1, Hongo, Bunkyo-ku, Tokyo, 113-0033, Japan}
\begin{abstract}
Many papers proved the security of quantum key distribution (QKD) system,
in the asymptotic framework.
The degree of the security 
has not been discussed in the finite coding-length framework, sufficiently.
However, to guarantee any implemented QKD system requires, 
it is needed to evaluate a protocol with a finite coding-length.
For this purpose,
we derive a tight upper bound of the eavesdropper's information.
This bound is better than existing bounds.
We also obtain the exponential rate of the eavesdropper's information.
Further, we approximate our bound by using the normal distribution.
\end{abstract}
\maketitle

\section{Introduction}
Quantum key distribution (QKD) was proposed by 
Bennett \& Brassard \cite{bene} as a 
protocol (BB84 protocol)
sharing secret keys by using quantum communication channel.
Their original protocol assumes a noiseless quantum channel,
but any quantum channel has noise in the realistic case.
Hence, the security of the BB84 protocol in this realistic case
had been an open problem for a long time, 
and has been proved by Mayers \cite{mayer1}.
He showed that the protocol becomes secure 
when the protocol is constructed by 
combining classical error correction and randomly choosing 
a code for privacy amplification.
In his proof, the secure generation key rate 
is $1-h(2p)-h(p)$ where $p$ is the qubit error rate
and $h(p)$ is the binary entropy $-p\log p -(1-p)\log (1-p)$
and the base of the logarithm is 2.
He also gave a bound of Eve's information
for a finite length code.
His discussion was extended to a more realistic framework by 
Inamori, L\"{u}tkenhaus \& Mayers \cite{ILM}.

After Mayers' proof, Shor \& Preskill \cite{shor2}
proved the security based on the method of 
Calderbank-Shor-Steane (CSS) codes\cite{C-S,Steane}.
Then, they proved the existence
of code achieving the secure generation key rate $1-2 h(2p)$ 
and pointed the possibility of 
the secure generation key rate $1-2 h(p)$. 
After their discussion,
treating the reliability of CSS codes,
Hamada\cite{Hamada03} showed the existence of the code 
attaining the secure generation key rate $1-2 h(p)$.
He also derived a bound of Eve's information
for a finite length code, which yields
the asymptotic secure generation key rate $1-2 h(p)$.
However, they did not discuss the complexity of 
the encoding and decoding\cite{shor2,Hamada03},
while the complexity of privacy amplification
is not so large in Mayers proof\cite{mayer1}.

Following these researches, 
Christandl, Renner, \& Ekert \cite{CRE}, 
Renner, Gisin, \& Kraus \cite{RGK}, and   
Koashi \cite{Koashi} showed
that the asymptotic secure generation key rate $1-2 h(p)$ is attained
when the protocol is constructed by 
combining classical error correction and randomly choosing.
However, they did not give the 
bound of Eve's information of the finite-length code, explicitly.
S. Watanabe, R. Matsumoto \& Uyematsu \cite{WMU}
considered Eve's information for a finite length code 
based on random privacy amplification, which yields
the asymptotic secure generation key rate $1-2 h(p)$.

On the other hand, Stucki {\em et al.} \cite{SGGRZ}
demonstrated quantum key distribution over 67 km between Geneva and Lausanne.
Kimura {\em et al.} \cite{KNHTKN04}
succeeded 150 km QKD transmission
with the error rate 8 -- 9 \%.
Also Gobby {\em et al.} \cite{GYS} did 122 km QKD transmission with 
the error rate 8.9 \%.
Tanaka {\em et al} \cite{TMTT05} demonstrated 
a continuous quantum key distribution 
over 16.3 km commercial use fiber with 14 days, and Yuan \& Shields \cite{Y-S}
did it over 20.3 km installed telecom fiber with 19 hours.
In these experiments, 
they succeeded in realizing the real system that could become truly secure 
if the coding system with infinite coding-length.
Hence, there is no implemented system whose security is guaranteed.
Thus, it is required to realize the error correcting code and 
the privacy amplification for guaranteeing the security of the 
implemented QKD system.

However, the required sizes of the error correcting code 
and the random privacy amplification 
are not clarified for a given quantum bit error rate, {\em e.g.}, 8 \%.
Therefore, 
many QKD experimental researchers 
want to know 
a tighter upper bound of Eve's information 
for given sizes of 
the classical error correcting code and the random privacy amplification.

In this paper, we derive an upper bound of 
Eve's information satisfying the following conditions,
first time.
1) The upper bound depends only on the size of
random privacy amplification.
2) By using this bound, the key generation rate $1-2h(p)$
can be attained.
In fact, Mayers' discussion \cite{mayer1} gives the upper bound in 
the finite-length case, but
his discussion yields the rate 
$1-h(2p)-h(p)$ not the rate $1-2h(p)$.
The discussion by S. Watanabe et al. \cite{WMU}
yields the rate $1-h(p)$,
but the bound depends on the error correction.
Koashi's discussion \cite{Koashi} satisfies the conditions 1) and 2),
but his discussion
does not clearly give the bound in the 
finite-length case.
Further, the protocol in his paper \cite{Koashi} 
and his older paper \cite{K-P}
is slightly different from the simple combination of 
the classical error correction and the random privacy amplification.
Our upper bound is also better than that by S. Watanabe et al. \cite{WMU}.

Moreover, it is shown that our evaluation cannot be further improved 
in the sense of the exponential rate
when the classical error correcting code satisfies a specific condition.
In this case, the exponential rate of our upper bound of 
Eve's information can be attained by a collective attack, 
which is realized by individual operation to the channel and 
the collective operation to Eve's local memory,
while our bound is valid even for the coherent attack,
which includes any Eve's attacks allowed by the physical principle.
That is, any coherent attack cannot improve
the best collective attack in the sense of the exponential rate
of Eve's information.
Indeed, Renner et al. \cite{RGK} proved that 
it is sufficient to show the security for
collective attacks for the treatment of the asymptotic key generation
rate since any channel can be approximated by 
a separable channel by using random permutation.
This result can be regarded as the extension of Renner's result 
to the exponential framework.
Also, this implies that our evaluation gives the optimal (minimum) 
exponential rate of Eve's information.

There is another type of asymptotic treatment 
else the exponential treatment.
In statistics, 
when the variable obeys the independent and identical distribution,
its distribution can be approximated by the normal distribution.
We also succeeded in approximating our upper bound by 
using the normal distribution.
In this approximation, we treat the asymptotic behavior 
when the size of the random privacy amplification is given as the form 
$2^{n h(\hat{p}_\times + \epsilon(\hat{p}_\times))}$
for the estimate $\hat{p}_\times$ of the phase error rate
while in the large deviation case (the exponential rate case)
we treat it when the size is given as the form 
$2^{n h(\hat{p}_\times + \tilde{\epsilon}(\hat{p}_\times)/\sqrt{n})}$,
where $\epsilon$ and $\tilde{\epsilon}$ are functions of $\hat{p}_\times$.

Here, we should remark that our results cannot be obtained 
by the combination of existing results.
The main technical point is
the relation between Eve's information and the phase error probability,
which is given in Lemma \ref{le-1}.
Owing to this lemma,
Eve's information can be bounded without 
any discussion of classical error correcting code for bit error.
Further, in association with the error correction of phase error,
we obtain an upper bound of the average error probability of 
a modified random coding 
when minimum Hamming distance decoding is applied (Lemma \ref{le-2}).
Combining these new techniques,
we obtain the upper bounds (Theorems \ref{th-1} and \ref{th-5})
through a long careful derivation.

In the following, the organization of this paper is explained.
First, we briefly explain classical error correcting code,
and describe our protocol using this knowledge
in section \ref{s2}.
In section \ref{s3}, we give an upper bound of 
Eve's information per one code and 
that of Eve's information per one bit.
The random privacy amplification corresponds to 
the random coding concerning the phase error.
Hence, we treat the average error of random coding
in section \ref{s4}.
Generalized Pauli channel is known as an important class of 
noisy channels. 
In quantum key distribution,
the noisy channel does not necessarily 
belong to this class.
However, if we use linear codes, 
we can treat any noisy channel as 
a generalized Pauli channel.
We summarize notations and properties of 
generalized Pauli channel in section \ref{s5}.
In section \ref{s6}, we prove the main theorem by assuming 
a upper bound of Eve's information when 
Eve's attack is known.
In section \ref{s7}, 
we derive a relation between the phase error and 
Eve's information.
In section \ref{s9},
the bound used in section \ref{s6} is proved by using 
the properties of generalized Pauli channel,
the bound of average error and 
the relation obtained in section \ref{s7}.

Further, we give the asymptotic behavior 
in the two asymptotic frameworks in section \ref{s3}.
Asymptotic formulas for 
large deviation and 
limiting distribution
are proved in Appendixes \ref{a2} and \ref{a3}, respectively.
Based on this evaluation, we compare our large deviation bound 
with
the bound by S. Watanabe, R. Matsumoto \& Uyematsu \cite{WMU}.
Further, in section \ref{s10}, 
we prove that the exponential rate of our bound of Eve's information
can be attained by the collective attack
under a specific condition.

\section{Protocol}\label{s2}
In this section, we describe our protocol.
Since our protocol employs the method of 
classical error correcting code,
we first explain classical error correcting code
for the preparation of description of our protocol.
\subsection{Classical error correcting code}
When the noise in a binary signal $\bF_2=\{0,1\}$ is symmetry,
the binary channel is described by 
a probability distribution $\{p,1-p\}$.
In this case, when we send a binary string (in $\bF_2^n$),
the noise can be described by a binary string $N$
and is characterized by the distribution $P$ on $\bF_2^n$.
Then, when the input signal is described by
the random variable $X$,
the output signal is described by the random variable $X+N$.
Error correcting code is a method removing the difference $N$.
In an error correcting code with $n$-bit,
we prepare an $m$-dimensional linear subspace $C$ of $\bF_2^n$, and 
the sender (Alice) and the receiver (Bob) agree that 
only elements of $C$ is sent before the communication.
This linear subspace is called a code or a $[n,m]$ code.
In this case, 
an encoding is given by a linear map $G(C)$ from $\bF_2^m$ to $C$.
Of course, the map $G(C)$ is given as 
an $m \times n$ matrix with $0,1$ entries.
Hence, when Bob receives an element out of $C$, 
he can find that there exists a noise,
and choose the most proper element among $C$ based on 
the obtained binary string.
Here, we can correct only one element among 
each equivalent class $[X] \in \bF_2^n/C$.
More precisely, we choose 
the most likely noise $\Gamma([X])$ among each equivalent class $[X]$.
This element is often called the representative,
and the set of representatives is denoted by $\Gamma$.
More generally, the decoding process is described by 
a map $D:\bF_2^n \to \bF_2^m$.

Hence, when Bob receives $X+N$, he decode it to 
$X+N-\Gamma([X+N])=X+ N-\Gamma([N])$.
Thus, the decoding error is described by the behavior of the 
random variable $N-\Gamma([N])$,
and does not depend on the input signal $X$.
When the noise belongs to the set $\Gamma$,
we can properly correct the error.
The error probability is equal to $1-P(\Gamma)$.

Suppose that there exists an eavesdropper (Eve) obtaining 
some information concerning the original signal $X$.
In this case, we prepare a linear subspace $C'$ of $C$
and Alice sends the information as 
an element of $C/C'$.
That is, when he sends an information corresponding to 
$[X] \in C/C'$,
he chooses one element among $[X]$ with the equal probability,
and sends it.
This operation is called 
privacy amplification.

\subsection{Our protocol}
Using this method,
we can reduce Eve's information.
However, it is not easy to evaluate how much information 
Eve has in this case.
The purpose of this paper is 
evaluating Eve's information.
In this case, 
the probability that 
Bob recovers the original information 
correctly is equal to P$(\Gamma + C')$, where
$\Gamma + C':=\{ \Gamma([X])+ X'| X\in \bF_2^n,X'\in C'\}$.
In addition, when we choose each linear subspace $C'$ of $C$
with the equal probability and we regard $C$ as $\bF_2^m$ and 
$C/C'$ as $\bF_2^{m-\tilde{m}}$,
the function from $\bF_2^m$ to $\bF_2^{m-\tilde{m}}$ 
is called the universal hashing function.
This function is can be constructed as 
an $(m-\tilde{m}) \times m $ matrix by choosing the elements with 
the uniform distribution.

Using this preparation, we briefly describe our protocol for 
quantum key distribution that can be realized by small complexity.
After this description, we present it precisely.
In our protocol, after quantum communication,
Alice and Bob check their basis by using public channel,
announce a part of obtained bits,
and estimate the bit error rate $p_+$ and phase error rate $p_\times$.
Here, we denote Alice's remaining bit string with 
the $+$ basis and the $\times $ basis
by $X_+$ and $X_\times$, respectively.
Similarly, we denote Bob's remaining bit string 
by $\tilde{X}_+$ and $\tilde{X}_\times$.
These bit strings are called raw keys.
Hence, the rates of $1$ in the difference $N_+= X_+ - \tilde{X}_+$ and 
the difference $N_\times= X_\times - \tilde{X}_\times$
are almost equal to $p_+$ and  $p_\times$, respectively.

Using the following process, Alice and Bob remove their errors and 
share the bit string with almost no error.
Alice generates another bit string $X'$
and sends the bit string $K:=X'+X_+$ to Bob.
Based on the information $K$, Bob 
obtains the information $X'':=K-\tilde{X}_+= X'+N_+$.
Using this method, we can realize a classical channel 
with the input $X'$ and the output $X''$.
The error rate of this channel is almost equal to $p_+$.
By applying classical error correction to this channel,
Alice and Bob can share bit string with almost zero error.
In this case, Alice generates an element
$X' \in \bF_2^m \cong C$,
and Bob recovers $X'''=D(X'')$.
Then, $X'''$ coincides with $X'$ in a high probability.
Finally, Alice and Bob perform the above mentioned hashing function
for their respective keys.
That is, Alice generates the $(m-l) \times m$ matrix $A$
with the rank $m-l$ randomly, and send this matrix.
Then, Alice and Bob 
obtain their final keys
$A X'$ and $A X'''$.

Therefore, the rate of final key to the raw key
is equal to $R=\frac{m-l}{n}$.
Roughly speaking, it is suitable to choose 
$m$ as an integer a little smaller than $(1-h(p_+))n$, and 
$\tilde{m}$ as an integer a little larger than $h(p_\times)n$.
Then, the generation rate $R$ is almost equal to 
$1-h(p_+)-h(p_\times)$.

In the following, 
we describe our protocol more precisely.
For this purpose, we need some mathematical notations.
The quantum system of each quantum signal is 
the two-dimensional Hilbert space $\cH_2$,
which is spanned by the $\{ | a\rangle \}_{a \in \bF_2}$.
We need to fix
integers $n_+$, $l_+$, $m_+$, $n_{\times}$, $l_{\times}$,
and $m_{\times}$
that describe the size of our code.
For a classical error correction, 
we choose an $m_+$-dimensional classical code $C_{1,+}$ in $\bF_2^{n_+}$ 
(an $m$-dimensional linear space $C_{1,+}$ of $\bF_2^{n_+}$),
and an $m_{\times}$-dimensional classical code 
$C_{1,\times}$ in $\bF_2^{n_\times}$.
We also fix the thresholds $\underline{k}_+$, $\overline{k}_+$, 
$\underline{k}_\times$, and $\overline{k}_\times$,
and the allowable statistical fluctuation $\delta_k$ for each 
count $k$ of error.

\begin{enumerate}
\item[(i)]
The sender, Alice, and the receiver, Bob, repeat steps 
(ii)--(iv) for each $i$.
\item[(ii)]Alice chooses a random bit $\ba_i$ and 
a random bit $\bb_i$.
\item[(iii)]
Bob chooses a random bit $\bc_i$.
\item[(iv)]
When $\bb_i = 0$,
Alice sends the quantum state $| \ba_i \rangle$,
otherwise, does the state 
$\frac{1}{\sqrt{2}}
(|0\rangle + (-1)^{\ba_i }| 1\rangle$.
In the following, $\{|0\rangle, |1\rangle\}$
is called the $+$ basis, and
$\{
\frac{1}{\sqrt{2}}
(|0\rangle + | 1\rangle),
\frac{1}{\sqrt{2}}(|0\rangle - |1\rangle)\}$
is called the $\times$ basis.
\item[(v)]
Alice and Bob announce $\bb_i$ and $\bc_i$ 
and discard any results for $\bb_i \neq\bc_i$.
They obtain 
$n_+ + l_+$ bits sequence with $\bb_i =\bc_i=0$,
and 
$n_\times + l_\times$ bits sequence with $\bb_i =\bc_i=1$.
\item[(vi)]
Alice randomly chooses $l_+$ check bits 
$X_{+,c,1},\ldots,X_{+,c,l_+}$
among $n_+ + l_+$ bits with the $+$ basis
and $l_\times$ check bits $X_{\times,c,1},\ldots,X_{\times,c,l_+}$
among $n_\times + l_\times$ bits with the $+$ basis, 
announces the positions of these bits,
and sends their information.
They obtain 
the estimates $\hat{p}_ +$ and $\hat{p}_ \times$ 
with the respective basis.
That is,
they count the number of error bits 
$k_+=|\{i|X_{+,c,i}\neq \tilde{X}_{+,c,i}\}|$ and 
$k_\times=
|\{i|X_{\times,c,i}\neq \tilde{X}_{\times,c,i}\}|$,
where $\tilde{X}_{+,c,i}$ and 
$\tilde{X}_{\times,c,i}$ are Bob's check bits.
However, when $k_+$ is greater than the threshold $\overline{k}_+$,
they discard their remaining bits with the $\times$ basis.
When $k_\times$ is greater than the threshold $\overline{k}_\times$,
they discard their remaining bits with the $+$ basis.
Further, when $k_+$ is less than the other threshold $\underline{k}_+$,
they replace $k_+$ by $\underline{k}_+$.
When $k_\times$ is less than the other threshold $\underline{k}_\times$,
they replace $k_\times$ by $\underline{k}_\times$.
\end{enumerate}
In the following, we treat only the bit string 
of the $+$ basis.
We denote Alice's (Bob's) remaining $n_+$-bit strings with the $+$ basis
by $X_+$ ($\tilde{X}_+$).
After this process, they apply the same procedure to 
the remaining bit strings with the $\times$ basis.
\begin{enumerate}
\item[(vii)]
Alice generates $Z_{+} \in \bF_2^{m_+}$
randomly, and sends Bob $G(C_{1,+} )Z_{+} + X_{+}$.
\item[(viii)]
Bob obtains the signal
$G(C_{1,+} )Z_{+}  + X_{+} -\tilde{X}_{+} \in \bF_2^{n_+}$.
Performing the decoding of the code $C_{1,+}  \approx \bF_2^{m_+}$,
he obtains $\tilde{Z}_{+} \in \bF_2^{m_+}$.
\item[(ix)]
Alice chooses 
$\tilde{m}:= n_{\times} 
h(k_\times/l_\times + \delta_{k_\times})$-dimensional subcode 
$C_{2, + } (Y_ +  ,k_\times) 
\subset \bF_2^{m_+}$
based on random variables $Y_+$ 
such that
any element $x \neq  0 \in \bF_2^{m_+}$
belongs to $C_{2, + } (Y_ +  ,k_\times) $ with the probability
$\frac{2^{n_+ h(k_\times/l_\times + \delta_{k_\times})}-1}{2^{m_+}-1}$.
\item[(x)]
Alice obtains the secret information
$\overline {Z}_+ 
:= \left[ Z_ + \right]_{C_{2, + } (Y_ +  ,k_ \times)}  
\in \bF_2^{m_+} /C_{2, + } (Y_ +  ,k_\times)$. 
\item[(xi)]
Bob obtains the secret information 
$ \overline {Z}_{+,B}   
:= \left[ \tilde{Z}_ + \right]_{C_{2, + } (Y_ +  ,k_\times)}  
\in \bF_2^{m_+} /C_{2, + } (Y_ +  ,k_ \times)$. 
\end{enumerate}

For example, $s$-dimensional code $C_2(Y,s)$ in $\bF_2^{m_+}$
is constructed based on $k$ random variables
$Y:= (X_1 , \cdots ,X_s)$  in $\bF_2^{m_+}$
as
$C_2(Y,s): = \left\langle X_1 , \cdots ,X_s  \right\rangle$,
where $Y$ obeys the uniform distribution
on the set $\{Y|X_1 , \cdots ,X_s\hbox { are linearly independent.}\}$.

\subsection{Extension of our protocol}
Indeed, in the realistic case, 
the bottleneck is often
the estimation error of the error rate.
Hence, in order to decrease the error of the estimation of 
the phase error rate $p_\times$,
we propose the following the modified protocol for any integer $a$.
In the modified protocol,
we repace steps (v) and (vi) by the following,
and add the step (xii).

\begin{enumerate}
\item[(v)]
Alice and Bob announce $\bb_i$ and $\bc_i$ 
and discard any results for $\bb_i \neq\bc_i$.
They obtain 
$a n_+ + l_+$ bits sequence with $\bb_i =\bc_i=0$,
and 
$a n_\times + l_\times$ bits sequence with $\bb_i =\bc_i=1$.

\item[(vi)]
Alice randomly chooses $n_+$ bits among remaining $an_+$ bits
with $+$ basis
and obtain $n_+$ bit string $X_{+}$.
She also sends the her positions to Bob.
Bob obtains the $n_+$ bit string $\tilde{X}_+$. 
They do the same procedure for the $\times$ basis.

\item[(xii)]
They repeat steps (vii) -- (xi) $a$ times.
\end{enumerate}

In the above protocol, 
the estimation of the phase error $p_\times$ 
has the same accuracy as
that of the first protocol 
with $a l_\times$ check bits of the $\times$ basis.

\section{Security}\label{s3}
In this section, we evaluate the security of our protocol.
In the following, for simplicity, we abbreviate
$l_\times$ and $n_+$ by
$l$ and $n$, respectively.
\subsection{Finite-length case}
The security of this protocol is evaluated by the 
mutual information $I(\overline{Z}_+,Z_E) $ between 
Alice's final key $\overline{Z}_+$ and eavesdropper(Eve)'s information
$Z_E$.
It is mathematically defined by
\begin{align*}
I(\overline{Z}_+,Z_E):=&
-\sum_{Z_E }
P(Z_E)
\log P(Z_E) \\
&+\sum_{\overline{Z}_+}P(\overline{Z}_+)
\sum_{Z_E }
P(Z_E|\overline{Z}_+)\log P(Z_E|\overline{Z}_+).
\end{align*}
In order to evaluate this value, we have to 
treat the 
hypergeometric distribution
$P_{hg}(k|n,l,j):=
\frac{
\genfrac{(}{)}{0pt}{}{l}{k}
\genfrac{(}{)}{0pt}{}{n}{j-k}
}{
\genfrac{(}{)}{0pt}{}{n+l}{j}}$.
This is because the random sampling 
obeys the hypergeometric distribution.
It is known that
its average is $\frac{lj}{n+l}$
and its variance is
$\frac{j l n (n+l-j)}{(n+l)^2 (n+l-1)}$.
In this paper, 
we focus on 
the average of Eve's information
$\rE_{\pos_\times,k_\times,Y_ +|
\pos_+,k_+,Y_\times} 
[I\left( \overline{Z}_+ ,Z_E \right)] $
for each $n,l$,
where $\pos_+$ and $\pos_\times$ are the random variables indicating 
the positions of the check bit of $\times$ basis and $+$ basis,
respectively.
Some papers \cite{K-P,Koashi,mayer1,shor2,WMU}
guarantee the security by proving
that
for any $\epsilon_1 >0$ and $\epsilon_2 >0$
there exist integers $n$ and $l$ such that
\begin{align}
P (I\left( \overline{Z}_+ ,Z_E \right) \ge \epsilon_2)
\le \epsilon_1.\label{2-3-1}
\end{align}
Indeed, when
$\rE_{\pos_\times,k_\times,Y_ +|
\pos_+,k_+,Y_\times} 
\left[I\left( \overline{Z}_+ ,Z_E \right) \right]
\le \epsilon_1 \epsilon_2$,
Markov inequality guarantees 
the inequality (\ref{2-3-1}).
Hence, 
we can recover 
the probabilistic behavior (\ref{2-3-1}) of Eve's information 
from the evaluation of 
the average of Eve's information.
Therefore, 
in this paper,
we concentrate the evaluation of 
the average of Eve's information.

\begin{thm}\label{th-1}
When $R$ is the rate of the code $C_1$
and the threshold $\overline{k}$ is less than $\frac{n}{2}$, 
we have
\begin{align}
\rE_{\pos_\times,k_\times,Y_ +|
\pos_+,k_+,Y_\times} 
\left[
I\left( \overline{Z}_+ ,Z_E \right) \right]
\le 
P(\delta,n,l,\underline{k},\overline{k}),\label{2-2-30}
\end{align}
where
\begin{widetext}
\begin{align*}
P(\delta,n,l,\underline{k},\overline{k})
:=&
\max_j 
\bar{h}
\Bigl(
\sum_{k=0}^{\underline{k}}
P_{hg}(k|n,l,j)
f(j-\underline{k},\underline{k}|n,l,\delta_{k})
+
\sum_{k=\underline{k}+1}^{\overline{k}}
P_{hg}(k|n,l,j)
f(j-k,k|n,l,\delta_{k})
\Bigr)\\
&+
\max_j \Bigl[\sum_{k=0}^{\underline{k}}
P_{hg}(k|n,l,j)
f(j-\underline{k},\underline{k}|n,l,\delta_{k})
n(R-h(\underline{k}/l + \delta_{k}))
\\
&\hspace{9ex}+
\sum_{k=\underline{k}+1}^{\overline{k}}
P_{hg}(k|n,l,j)
f(j-k,k|n,l,\delta_{k})
n(R-h(k/l+ \delta_{k}))
\Bigr], \\
\end{align*}
\end{widetext}
and 
\begin{align*}
\bar{h}(x):=&
\left\{
\begin{array}{ll}
h(x) & x < 1/2 \\
1 & x \ge 1/2
\end{array}
\right.\\
f(k',k|n,l,\delta):=&
\left\{
\begin{array}{ll}
\min\{
2^{n(h(\frac{k'}{n})-h(\frac{k}{l}+\delta))}
,1\}& \hbox{ if } k' < n/2 \\
1 & \hbox{ if } k' \ge n/2 .
\end{array}
\right.
\end{align*}
\end{thm}
Further, Eve's information per one bit is evaluated as follows.
\begin{thm}\label{th-5}
When $R$ is the rate of the code $C_1$, 
we have
\begin{align*}
& \rE_{\pos_\times,k_\times,Y_ +|
\pos_+,k_+,Y_\times} 
\left[\frac{I\left( \overline{Z}_+ ,Z_E \right)}
{n(R-h(k_\times/l_\times +\delta_{k_\times}))} \right]\\
\le &\tilde{P}(\delta,n,l,\underline{k},\overline{k}),
\end{align*}
where
\begin{widetext}
\begin{align*}
& \tilde{P}(\delta,n,l,\underline{k},\overline{k}) \\
:= &
\max_j 
\frac{1}{n(R-h(\overline{k}/l +\delta_{\overline{k}}))}
\bar{h}
\Bigl(
\sum_{k=0}^{\underline{k}}
P_{hg}(k|n,l,j)f(j-\underline{k},\underline{k}|n,l,\delta_{k})
+
\sum_{k=\underline{k}+1}^{\overline{k}}
P_{hg}(k|n,l,j)
f(j-k,k|n,l,\delta_{k})
\Bigr)\\
& + \max_j \Bigl[
\sum_{k=0}^{\underline{k}}
P_{hg}(k|n,l,j)
f(j-\underline{k},\underline{k}|n,l,\delta_{k})
+\sum_{k=\underline{k}+1}^{\overline{k}}
P_{hg}(k|n,l,j)
f(j-k,k|n,l,\delta_{k})
\Bigr].
\end{align*}
\end{widetext}
\end{thm}

The proofs of these theorems are divided into two parts:
(i)The security of known channel (section VIII),
(ii)The security of unknown channel, which is given by 
estimating the channel and employing the part (i) (section VI).
For treatment of quantum channel, we prepare the notations of
generalized Pauli channel in section V.
For the discussion of the part (i), 
we derive a bound of 
average error concerning classical error correcting code in Section IV,
and a bound of Eve's information using the phase error in Section VII.

\subsection{Approximation using normal distribution}
In the following,
we calculate the above value approximately.
For this purpose, 
we choose two probabilities $\underline{p} < \overline{p}<\frac{1}{2}$,
and a continuous function $p \mapsto \tilde{\epsilon}(p)$.
When 
$\overline{k}=\overline{p}l$,
$\underline{k}=\underline{p}l$,
$\frac{n}{n+l}=r$,
$\delta_k=\frac{\tilde{\epsilon}(p)}{\sqrt{n+l}}$,
as is shown in Appendix \ref{a2},
we obtain
\begin{align}
\lim_{n \to \infty}
\tilde{P}(\delta,n,l,\underline{k},\overline{k})
=
\max_{p\in[\underline{p},\overline{p}]}
\Phi\left(-
\frac{\sqrt{r(1-r)}}{\sqrt{p(1-p)}}
\tilde{\epsilon}(p)
\right),\label{1-29-1}
\end{align}
where
the distribution function $\Phi$ of the standard Gaussian distribution:
\begin{align*}
\Phi(x):=
\int_{-\infty}^x
\frac{1}{2\pi}e^{-x^2/2}d x.
\end{align*}
Hence, 
in order to keep the security level $\varepsilon$ per one bit,
it is suitable to choose 
$\delta_k$ to be
$
-\frac{1}{\sqrt{n+l}}
\frac{\sqrt{\frac{k}{l}(1-\frac{k}{l})}}
{\sqrt{\frac{n}{n+l}\frac{l}{n+l}}}
\Phi^{-1}(\varepsilon)
=-\sqrt{\frac{n+l}{nl}}
\sqrt{\frac{k}{l}(1-\frac{k}{l})}
\Phi^{-1}(\varepsilon)$
when 
$\tilde{P}(\delta,n,l,\underline{k},\overline{k})$ can be approximated by 
the RHS of (\ref{1-29-1}).
That is, 
our upper bound is almost determined by 
$
\frac{\sqrt{\frac{nl}{n+l}}}
{\sqrt{\frac{k}{l}(1-\frac{k}{l})}}\delta_k$.

Now, we consider the case when 
we use a low-density-parity-check (LDPC) code as the code $C_1$ \cite{YMI}.
In this case,
the case of $R=0.5$, and $n=10,000$ is one realistic case.
As an realistic case, 
let us consider the case
$l=1,000$, $\overline{p}= 0.075$, $\delta_{k_\times}=0.01$.
Then, we have$
-
\frac{\sqrt{\frac{nl}{n+l}}}
{\sqrt{\frac{k}{l}(1-\frac{k}{l})}}\delta_k
= -1.14$.
The security 
level $\Phi\bigl(
-
\frac{\sqrt{\frac{nl}{n+l}}}
{\sqrt{\frac{k}{l}(1-\frac{k}{l})}}\delta_k
\bigr)= 
0.126$ is not sufficient.

However, it is not easy to increase the size $n$.
Hence, we adopt the modified protocol.
In this case, we replace only $l$ by the following values.
In the case of $l=20,000$,
the security level is almost 0.001.

\par
\begin{widetext}
\begin{tabular}{|c|c|c|c|c|c|c|}
\hline
$l$& 1,0000 & 10,000 & 20,000& 30,000 & 40,000 & 50,000 \\
\hline
$ 
-\frac{\sqrt{\frac{nl}{n+l}}}
{\sqrt{\frac{k}{l}(1-\frac{k}{l})}}\delta_k
$&
$-1.14$ & $-2.68$ & $-3.10$ & $-3.29$ & $-4.00$ & $-3.47$ \\
\hline
$\Phi\bigl(
-\frac{\sqrt{\frac{nl}{n+l}}}
{\sqrt{\frac{k}{l}(1-\frac{k}{l})}}\delta_k
\bigr)$
& 0.126 & 
0.00363 & 
0.000968 &
0.000505 &
0.000342 &
0.000264 \\
\hline
\end{tabular}
\par
\vspace{2ex}
\end{widetext}

\subsection{Large deviation}
Next, we focus on the large deviation type evaluation.
Choose a function $p \in [\underline{p},\overline{p}]
\mapsto \epsilon(p)$
and define
\begin{align*}
E(\epsilon,r,\underline{p},\overline{p})
:=&
\min_{p \in [\underline{p},\overline{p}],\epsilon' \ge 0}
\Bigl[
h(p + r (\epsilon(p)-\epsilon'))
\\
&
-(1-r)h(p)-2 r h(p + \epsilon(p)-\epsilon') \\
&+ r h( p +\epsilon(p))\Bigr].
\end{align*}
When $\overline{k}=\overline{p}l$,
$r=\frac{n}{n+l}$, $\delta_k= \epsilon(\frac{k}{l})$,
as is shown in Appendix \ref{a3},
we obtain
\begin{align}
E(\epsilon,r,\underline{p},\overline{p})
=\lim_{n \to \infty} 
\frac{-r}{n}\log P(\delta,n,l,\underline{k},\overline{k}).
\label{1-29-2}
\end{align}
Further,
\begin{align}
P(\delta,n,l,\underline{k},\overline{k})
\le &
\overline{k}(n+l+1)
n (R-h(\underline{p}+\delta_{\underline{p}}))
2^{\frac{-n}{r}E(\epsilon,r,\underline{p},\overline{p})} \nonumber \\
& \hspace{5ex} +
h(\overline{k}(n+l+1)
2^{\frac{-n}{r}E(\epsilon,r,\underline{p},\overline{p})}).
\label{1-29-3}
\end{align}
Hence, given a fixed real number $E$,
it is suitable to choose $\epsilon(p)$ satisfying that
\begin{align*}
E=&
\min_{\epsilon' \ge 0}\bigl[
h(p + r (\epsilon(p)-\epsilon'))-(1-r)h(p)\\
&-2 r h(p + \epsilon(p)-\epsilon')
+r h( p +\epsilon(p))
\bigr]
\end{align*}
for any probability $p \in [\underline{p},\overline{p}]$.
Further, when $\epsilon(p)$ is sufficiently small,
using
the relation
$d(p\|q):= 
p \log \frac{p }{q}+(1-p) \log \frac{1-p }{1-q}
\cong\frac{(p-q)^2}{p(1-p)\ln 2}
$,
we have the approximation.
\begin{align*}
& h(p + r (\epsilon(p)-\epsilon'))-(1-r)h(p)-r h(p + (\epsilon(p)-\epsilon')) \\
&+ r h( p +\epsilon(p))-r h(p + (\epsilon(p)-\epsilon'))\\
= &
(1-r) d(p \|p + r (\epsilon(p)-\epsilon') )
+ r d(p+\epsilon(p)  \| p + r (\epsilon(p)-\epsilon')) \\
&+ r (h( p +\epsilon(p))- h(p + (\epsilon(p)-\epsilon')))\\
\cong &
(1-r) \frac{r^2 (\epsilon(p)-\epsilon')^2}{p(1-p)}
+ r \frac{(1-r)^2 (\epsilon(p)-\epsilon')^2}{p(1-p)}
+ r h'(p) \epsilon'.
\end{align*}
In this approximation, 
when $\epsilon(p)$ is small enough,
the minimum is attained at $\epsilon'=0$.
Hence, 
\begin{align}
& \min_{\epsilon' \ge 0}\bigl[
h(p + r (\epsilon(p)-\epsilon'))-(1-r)h(p)-r h(p + \epsilon(p)-\epsilon')
\nonumber \\
&\hspace{5ex} + r h( p +\epsilon(p))-r h(p + \epsilon(p)-\epsilon') \bigr]
\nonumber \\
=&
h(p + r \epsilon(p))-(1-r)h(p)-r h(p + \epsilon(p))\label{2-2-22}.
\end{align}
The maximum value of $\epsilon(p)$ satisfying (\ref{2-2-22})
corresponds to the critical rate in the classical channel coding 
theory.
Therefore, when the number $\epsilon(p)$ is sufficiently small for each $p
\in [\underline{p},\overline{p}]$,
we obtain
\begin{align}
E&=
h(p + r \epsilon(p))-(1-r)h(p)-r h(p + \epsilon(p))\label{2-2-2}\\
&\cong \frac{r(1-r) \epsilon(p)^2}{(\log 2) (p + r \epsilon(p))(1-(p + r \epsilon(p)))},
\quad \forall p \in [\underline{p},\overline{p}].\nonumber
\end{align}
Hence, 
in this case, 
in order to keep the exponential rate $E$,
we choose $\epsilon(p)$ as
\begin{align*}
\epsilon(p)
=& \frac{
(\ln 2) E r (1-2p)}{2( r (1-r)+ (\ln 2) E r^2)} \\
& + \frac{\sqrt{(\ln 2)^2 E^2 r^2 + 4p(1-p)r (1-r) (\ln 2) E}}
{2( r (1-r)+ (\ln 2) E r^2)} \\
\cong &
\frac{\sqrt{p(1-p)}}{\sqrt{r(1-r)}}\sqrt{(\ln 2) E}
\hbox{  as  } E \to 0.
\end{align*}

Here, we compare our bound with 
that by S. Watanabe, R. Matsumoto \& Uyematsu \cite{WMU}.
Since their protocol is different from our protocol,
we compare our protocol with their protocol
with the same size of code.
This is because
the size of the code almost corresponds to 
the cost of its realization.
Then, their case corresponds to 
our case with 
$\overline{p}=\underline{p}=p$ and $l= n$.
They derived the following upper bound (\ref{6-22})
of the security in their protocol
when the codes $C_2 \subset C_1$ satisfy the following conditions:
The codes $C_1/C_2$ and $C_2^{\perp}/ C_1^{\perp}$
have the decoding error probability $\varepsilon$
when the channel is the binary symmetric channel with the error probability $p$.
\begin{align}
& \rE_{\pos_\times,k_\times|\pos_+,k_+} 
\left[I\left( \overline{Z}_+ ,Z_E \right)\right] \nonumber \\
\le &
h(
2(\frac{n}{2}+1)^2 \varepsilon
+
4(n+1)^2
e^{-\frac{\epsilon(p)^2}{4}n }
)\nonumber \\
&+
4n(\frac{n}{2}+1)^2 \varepsilon
+
8n(n+1)^2
e^{-\frac{\epsilon(p)^2}{4}n }.\label{6-22}
\end{align}
However, even if the error probability $\varepsilon$ is zero,
our evaluation (\ref{1-29-3}) is better than their evaluation (\ref{6-22}).
In particular, when 
$\epsilon(p)$ is sufficiently small,
we can use (\ref{2-2-22}).
From Pinsker inequality:
$(\ln 2) d(p\|q) \ge (p-q)^2$\cite{C-K},
our exponential rate is evaluated as
\begin{align*}
&\frac{\ln 2}{r}
(h(p + r \epsilon(p))-(1-r)h(p)-r h(p + \epsilon(p))) \\
=&
\frac{\ln 2}{r}
((1-r) d(p \|p + r \epsilon(p) )+ r d(p+\epsilon(p)  \| p + r \epsilon(p)))\\
\ge &
(1-r)\epsilon(p)^2 =\frac{\epsilon(p)^2}{2 },
\end{align*}
which is greater than their 
rate $\frac{\epsilon(p)^2}{4}$ even in the case of $\epsilon'=0$.
Further, our coefficient is smaller than their coefficient
in this case as follows:
\begin{align*}
& \overline{k}(n+l+1)
n (R-h(\underline{p}+\delta_{\underline{p}}))\\
\le &
p n (n+ n+1) nR 
< 8n(n+1)^2,\quad \\
& \overline{k}(n+l+1) 
=  p n (n+ n+1) <
4(n+1)^2
\end{align*}
because $p \le 1/2$.

Hence, in order to obtain a tighter bound, 
it is better to use our formula (\ref{2-2-30}).

\section{Error correcting code}\label{s4}
\subsection{Type method}
In this section, 
we treat classical error correcting code.
For this purpose, we review the type method for binary strings.
For any element $x \in \bF_2^n$,
we define $|x|:= | \{ i|x_i=1\}|$
and $T_n^k := \{ x \in \bF_2^n|~ |x|=k\}$.
Further, the number of elements
is evaluated by
\begin{align}
\frac{1}{n+1}2^{n h(k/n)}
\le |T_n^{k}|
= 
\genfrac{(}{)}{0pt}{}{n}{k}
\le
|\cup_{k'\le k}T_n^{k'}|
\le 2^{n h(k/n)}\label{1-16-30}
\end{align}
for $k \le n/2$.
For any distribution $P$ on $\bF_2^n$,
we define distribution $\tilde{P}$ on $\{0, \ldots, n\}$ 
and $P_k$ on $T_n^k$ as
\begin{align*}
\tilde{P}(k)&:= P(T_n^k) \\
P_k(x)      &:= 
\left\{
\begin{array}{ll}
\frac{P(x)}{\tilde{P}(k)}, & \hbox{ if } x \in T_n^k\\
0 &\hbox{ otherwise.} 
\end{array}
\right.
\end{align*}
Hence, we have
\begin{align*}
P(x)= \sum_{k=0}^n \tilde{P}(k) P_k(x)   .
\end{align*}

\subsection{Bound for random coding}
In this paper, we focus on linear codes,
which are defined as linear subspaces of $\bF_2^n$.
For the preoperation of the following section,
we consider the error probability when 
the noise of classical communication channel is given as
a classical channel $W$ (a stochastic transition matrix)
on $\bF_2^n$.
If a channel $W$ 
is written by a distribution $P_W$ on $\bF_2^n$ as
\begin{align*}
W(y|x)= P_W(y-x),
\end{align*}
it is called an additive channel.
For an additive channel $W$, we define the following distribution:
\begin{align*}
P_{W}(k):= P_W\{x| |x|= k \}.
\end{align*}
In order to protect our message from the noise,
we often restrict our message to be sent in a subset of $\bF_2^n$.
This subset is called a code.
When the noise is given by an additive channel,
a linear subspace $C$ of $\bF_2^n$ is suitable for our code
because of the symmetry of the noise.
Hence, in the following, we call a linear subspace $C$ of $\bF_2^n$ 
a code.

Now, for a preoperation of the following section,
we consider the error correcting code
using a pair of codes $C_1 \subset C_2$.
In order to send any information $[x_2]_1\in C_2/C_1$,
we send $x_1+x_2$ by choosing $x_1 \in C_1$ with 
the uniform distribution,
where $[x]_i$ denotes the equivalent class divided by $C_i$.
In this case, the decoder is described by 
the map $D$ from $\bF_2^n$ to itself.
When the channel is given by $W$,
the average error probability
is 
\begin{align*}
&P_{e,W}(D)\\
=&
\frac{1}{|C_2/C_1|}
\sum_{[x_2]_1\in C_2/C_1}
\frac{1}{|C_1|}
\sum_{x_1\in C_1}
\sum_{D(y)\neq [x_2]} W(y|x_2+x_1).
\end{align*}
However, 
we often describe our decoder by 
the coset representative $\Gamma([x]_2)$ for each 
$[x]_2\in \bF_2^n/C_2$.
That is, when the decoder receives the element $y$,
he decodes it to $D^{\Gamma}(y):=[y - \Gamma([y]_2)]_1$.
When the channel is given by a additive channel $W$,
the error probability is 
\begin{align*}
P_{e,W}(D^{\Gamma})=1-P_W (\Gamma+C_1),
\end{align*}
where $\Gamma:= \{\Gamma([x]_1)| [x]_1 \in  \bF_2^n/C_2$\},
and $\Gamma+C_1= \{ x+x_1| x \in \Gamma, x_1 \in C_1\}$.
For example, 
when we choose 
the minimum Hamming distance decoding 
$D_{C_2/C_1}$:
\begin{align*}
D_{C_2/C_1}(y):= 
\argmin_{[x_2]_1 \in C_2/C_1}\min_{x_1 \in C_1} |y-(x_1+x_2)|.
\end{align*}
By using the map $\Gamma([x]_2)$:
\begin{align*}
\Gamma([x]_2)=x+ \argmin_{x_2 \in C_2} |x+x_2|,
\end{align*}
it can be written as
\begin{align*}
D_{C_2/C_1}(y)= 
[y - \Gamma([y]_2)]_1.
\end{align*}
In the following, we denote the above $\Gamma$ by
$\Gamma_{C_2}$.

Now, we consider the average error when we choose the larger code $C_2$
randomly.
\begin{lem}\label{le-2}
Let $C_1$ be a arbitrary [n,t] code $(C_1 \subset \bF_2^n)$.
We randomly choose the $t+l$-dimensional code $C_2(X) \supset C_1$
such that any element $x \in \bF_2^n\setminus C_1$ belongs to
$C_2(X)$ with the probability $\frac{2^{l+t}-2^t}{2^n-2^t}$.
Then, any additive channel $W$ satisfies
\begin{align*}
&
\rE_X [P_{e,W} (D^{\Gamma_{C_2(X)}})]
=
\rE_X [1- P_{W} (\Gamma_{C_2(X)}+C_1)]
\\
\le &
\sum_{k=0}^{n}
\tilde{P}_{W}(k)
g(2^{l+t-n}|n,k),
\end{align*}
where
\begin{align*}
g(x|n,k):=
\left\{
\begin{array}{ll}
\min\{2^{n \overline{h}(k/n)}x,1\}& 
k \le \lfloor n/2 \rfloor\\
1 &
k > \lfloor n/2 \rfloor.
\end{array}
\right.
\end{align*}
\end{lem}
\begin{proof}
Let $T^n_k$ be the set $\{x\in \bF_2^n| |x|=k\}$.
Then,
$P(x)= \sum_{k=0}^n
\tilde{P}(k)P_k(x)
$.
Hence,
$P(\Gamma_{C_2(X)}+C_1)=
\sum_{k=0}^{n}
\tilde{P}(k)P_k(\Gamma_{C_2(X)}+C_1)$.

Indeed, 
if $y \in T^n_k \subset\bF_2^n$ 
does not belong to $\Gamma_{C_2(X)}+C_1$,
there exists an element 
$x \in C_2(X)\setminus C_1$ such that
$|y-x|\le k$.
Hence, the probability that 
at least one element 
belongs to the set 
$\{ x| |x-y |\le k\}$ 
is less than 
$2^{nh(k/n)}\frac{2^{l+t}-2^t}{2^n-2^t}$
for $k\le n/2$
because 
$|\{ x| |x-y |\le k\}|
=|\{ z| |z |\le k\}|
\le 2^{nh(k/n)}$.
(See (\ref{1-16-30}).)
Therefore,
\begin{align*}
& \rE_X 
[1- P_k(\Gamma_{C_2(X)}+C_1)] \\
\le & \sum_{y\in T_n^k}
P_k(y)2^{nh(k/n)}
\frac{2^{l+t}-2^t}{2^n-2^t} \\
\le & 2^{nh(k/n)}\frac{2^{l+t}-2^t}{2^n-2^t}
\le  2^{nh(k/n)}\frac{2^{l+t}}{2^n}
\end{align*}
for $k \le n/2$,
where 
the last inequality follows from $l+t \le n$.
This value is also bounded by $1$.
Hence,
\begin{align*}
& \rE_X [1- P(\Gamma_{C_2(X)}+C_1)] \\
=&
\sum_{k=0}^{n}
\tilde{P}(k)
\rE_X [1- P_k(\Gamma_{C_2(X)}+C_1)]\\
\le &
\sum_{k=0}^{n}
\tilde{P}(k)
g(2^{l+t-n}|n,k).
\end{align*}
\end{proof}

\section{Generalized Pauli channel}\label{s5}
In this section, 
for the preparation of our proof,
we give some notations concerning
generalized Pauli channels.
In order to describe it, 
for any two elements $x=(x_1,\ldots,x_n),
y=(y_1,\ldots,y_n)\in \bF_2^n$, we use the product:
\begin{align*}
x\cdot y := \sum_{i=1}^n x_i y_i.
\end{align*}
Thus, the space $\cH_2^{\otimes n}
= (\complex^2)^{\otimes n}$
is spanned by the $\{ | x\rangle \}_{x \in \bF_2^n}$.
Now, we define the unitary matrices
$\bX^x$ and $\bZ^z$ for $x,z \in \bF_2^n$ as:
\begin{align*}
\bX^x|x'\rangle &= |x'-x\rangle\\
\bZ^z|x'\rangle &= (-1)^{x'\cdot z} |x'\rangle.
\end{align*}
From the definition, we have the relation \cite{weyl}.
\begin{align*}
(\bX^x \bZ^z)(\bX^{x'} \bZ^{z'})
= (-1)^{x\cdot z' - x'\cdot z}
(\bX^{x'} \bZ^{z'})(\bX^x \bZ^z).
\end{align*}
When the channel $\Lambda$ has the form:
\begin{align*}
\Lambda(\rho)= \sum_{x,z\in\bF_2^n}
P_{\Lambda}(x,z)
(\bX^x \bZ^z)\rho(\bX^{x} \bZ^{z})^{\dagger},
\end{align*}
it is called a generalized Pauli channel.
Indeed, a generalized Pauli channel
is a quantum analogue of an additive channel.
In fact, 
it is known \cite{BDSW,Hama2}
that the channel $\Lambda$ is generalized Pauli
if and only if
\begin{align}
\Lambda(\rho)= 
(\bX^x \bZ^z)^{\dagger}
\Lambda((\bX^x \bZ^z)\rho(\bX^{x} \bZ^{z})^{\dagger})
(\bX^x \bZ^z),\quad
\forall x,z \in \bF_2^n.\label{1-4-1}
\end{align}
For any channel $\Lambda$,
we often focus on 
its twirling $\Lambda_t$ defined as
\begin{align*}
\Lambda_t(\rho)&:= 
\frac{1}{2^{2n}}\sum_{x,z \in \bF_2^n}\Lambda^{x,z}(\rho)\\
\Lambda^{x,z}(\rho)&:=
(\bX^x \bZ^z)^{\dagger}
\Lambda((\bX^x \bZ^z)\rho(\bX^{x} \bZ^{z})^{\dagger})
(\bX^x \bZ^z).
\end{align*}
From (\ref{1-4-1}), 
the twirling $\Lambda_t$ is always a generalized Pauli channel.

In the treatment of generalized Pauli channels,
the distribution $P_{\Lambda}(x,z)$ is important.
Hence, we introduce some notations for this distribution.
We define the 
distributions $P_{\Lambda,X}(x)$ and $P_{\Lambda,Z}(z)$ as
\begin{align*}
P_{\Lambda,X}(x)
:=\sum_{z \in \bF_2^n}P_{\Lambda}(x,z),\quad
P_{\Lambda,Z}(z)
:=\sum_{x \in \bF_2^n}P_{\Lambda}(x,z).
\end{align*}
These are called marginal distributions.
We also define the conditional distribution as
\begin{align*}
P_{\Lambda,Z|X}(z|x)
:=
\frac{P_{\Lambda}(x,z)}{P_{\Lambda,X}(x)}.
\end{align*}
Next, we treat a generalized Pauli channel $\Lambda$
on the tensor product system 
$(\complex^2)^{\otimes n_1}\otimes(\complex^2)^{\otimes n_2}$.
In this case, we use the following notation.
\begin{align}
P_{\Lambda,1}(x_1,z_1) &:= \sum_{x_2,z_2 \in \bF_2^{n_2}}
P_{\Lambda}(x_1x_2,z_1z_2 )\nonumber \\
P_{\Lambda,2}(x_2,z_2) &:= \sum_{x_1,z_1 \in \bF_2^{n_1}}
P_{\Lambda}(x_1x_2,z_1z_2 )\nonumber \\
P_{\Lambda,X,i}(x_i) &:= \sum_{z_i \in \bF_2^{n_i}}
P_{\Lambda,i}(x_i,z_i)\nonumber \\
P_{\Lambda,Z,i}(z_i) &:= \sum_{x_i \in \bF_2^{n_i}}
P_{\Lambda,i}(x_i,z_i)\nonumber \\
\tilde{P}_{\Lambda,Z,1,2}(k_1,k_2) &:= 
\sum_{x_i \in \bF_2^{n_i}}
P_{\Lambda,Z}(T_{n_1}^{k_1}\times T_{n_2}^{k_2})
\label{6-15}\\
P_{\Lambda,1|Z,2}(x_1,z_1|z_2)&:=
\frac{
\sum_{x_2\in \bF_2^{n_2}}
P_{\Lambda,1|Z,2}(x_1 x_2 ,z_1 z_2)}
{P_{\Lambda,Z,2}(z_2)} \nonumber \\
P_{\Lambda,Z,1|Z,2}(z_1|z_2)&:=
\sum_{x_1 \in \bF_2^{n_1}}
P_{\Lambda,1|Z,2}(x_1,z_1|z_2).\nonumber 
\end{align}
Note that $\tilde{P}_{\Lambda,Z,1,2}$ is different from 
$\tilde{P}_{\Lambda,Z}$.
These notations will be used in the following sections.

\section{Proof of Main Theorem}\label{s6}
\subsection{Modified protocol}
In this section, we prove 
Theorem \ref{th-1} by treating the security of the following protocol.
In the following protocol,
we fix the generalized Pauli channel $\Lambda$
from $n$-qubits system to itself.

\begin{enumerate}
\item[(i)]
Alice generates $Z_{+} \in \bF_2^m$
randomly, and sends Bob $G(C_1 )Z_{+} \in \bF_2^n$ with the $+$ basis
through the $n$-qubits generalized Pauli channel $\Lambda$.

\item[(ii)]
Bob measures the received $n$ qubits with the $+$ basis.
Performing the decoding of the code $C_1  \approx \bF_2^m$,
he obtains $\tilde{Z}_{+} \in \bF_2^m$.

\item[(iii)]
They do the processes (ix) -- (xi) of the previous protocol.
In this case, we assume that the dimension of 
the code $C_1$
(the subcode $C_{2,+}(Y_+)$)
is $t$ ($s$).
\end{enumerate}
This protocol is 
the special case that the channel is known.

For any channel $\Lambda$ from the system $\cH$ to itself,
the state on the environment system 
can be described 
by using its Stinespring representation $(\cH_E,U,|0\rangle_E\in \cH_E)$:
\begin{align*}
\Lambda(\rho)=
\Tr_{\cH_E} U \rho \otimes |0\rangle_E ~_E\langle 0|U^*.
\end{align*}
That is, the state on the environment system 
is characterized by another channel $\Lambda_E(\rho):=
\Tr_{\cH_2^n} U \rho \otimes |0\rangle_E ~_E\langle 0|U^*$.

In this above protocol,
the distribution of Eve's signal $Z_E$ 
is described by a POVM $M_{Z_E}$ on $\cH_E$
as $P(Z_E|Z)=\Tr M_{Z_E} \Lambda_E(\rho_{Z})$.
Therefore, 
in order to evaluate the classical mutual information
$I(\overline{Z},Z_{E})$
it is sufficient to evaluate 
the quantum mutual information (Holevo information)
\begin{align}
&I([z]\in C_1/C_2(Y),\rho_{\Lambda,E}^{C_1/C_2(Y)}([z]) ) \nonumber \\
:=& 
\frac{1}{2^{m-s}}
\sum_{[z]\in C_1/C_2(Y)} 
\Tr  \rho_{\Lambda,E}^{C_1/C_2(Y)} ([z])
\nonumber \\
& \hspace{10ex}
\cdot \Bigl( \log \rho_{\Lambda,E}^{C_1/C_2(Y)} ([z])
- \log \rho_{\Lambda,E} ^{C_1/C_2(Y)}
\Bigr) ,\label{2-5-2}
\end{align}
where
$\rho_{\Lambda,E}^{C_1/C_2(Y)} ([z])
:= \sum_{z_2\in C_2(Y)}
\Lambda _E (\left| z+z_2\right\rangle \left\langle z+z_2 \right|)$
and 
$\rho_{\Lambda,E} ^{C_1/C_2(Y)}:=
\frac{1}{2^{t-s}}
\sum_{[z]\in C_1/C_2(Y)} 
\rho_{\Lambda,E}^{C_1/C_2(Y)}([z])$.
In the following, we often 
abbreviate (\ref{2-5-2}) as
$I_H(\overline{Z},Z_{E})$.

\begin{thm}\label{th-2}
We can evaluate Eve's information as follows.
\begin{align*}
& \rE_{Y_+}
\left[I([z]\in C_1/C_2(Y_+,s),\rho_{\Lambda,E}^{C_1/C_2(Y)} ([z]))\right]
\nonumber \\
\le &
\eta_{m-s}
\Bigl(
\sum_{k=0}^{n}
\tilde{P}_{\Lambda,Z}(k)
g(2^{-s}|n,k)
\Bigr),
\end{align*}
where
$m=\dim C_1$ and 
$\eta_k$ is defined as
\begin{align*}
\eta_k(x):=\overline{h}(x)+k x.
\end{align*}
\end{thm}
This theorem will be proved in Section \ref{s9}.

\subsection{Proof of Theorem \ref{th-1}}
Now, we back to our main protocol.
First, we fix the random variables
$\pos_+, k_+,Y_\times$.
Then, it is sufficient to 
treat the quantum system of the $n_+ +l_{\times}$ qubits.
In the following, we characterize 
the system of raw keys $\complex^{n}$ by the subscript $k$,
and 
the other system of check qubits $\complex^{l_\times}$ by the subscript $c$.

Hence, we denote the quantum channel of this system by $\Lambda$.
Note that $\Lambda$ is not necessarily generalized Pauli.
In the following, we abbreviate 
$l_\times, \pos_\times,k_\times, Y_+$ 
by 
$l, \pos,k, Y$, respectively.

In this case, the variable $\pos$ takes a subset of 
$l$ elements $\{i_1, \ldots, i_l\} \subset \{1, \ldots, n+l\}$,
where $i_1< \ldots < i_l$.
Then, 
we define the unitary matrix $U_{\pos}$ as
\begin{align*}
U_{\pos}(u_{i_1}\otimes \cdots \otimes u_{i_l}
\otimes u_{j_1}\otimes \cdots \otimes u_{j_n})
=u_1 \otimes \cdots \otimes u_{n+l},
\end{align*}
where 
$\{j_1, \ldots, j_n\}= \{i_1, \ldots, i_l\}^c$
and $j_1 < \cdots < j_n$.
Every subset is choosed with the probability 
$\frac{1}{\choose{n+l}{l}}$.
We also define the channel $\Lambda^{\pos}$ for any channel $\Lambda$ as
\begin{align*}
\Lambda^{\pos}(\rho):=
U_{\pos}^\dagger (\Lambda (U_{\pos} \rho U_{\pos}^\dagger))U_{\pos}.
\end{align*}
Then, we can show that
\begin{align}
(\Lambda^{\pos})_t= (\Lambda_t)^{\pos}.\label{1-6-10}
\end{align}
Hence, any generalized Pauli channel $\Lambda$
satisfies 
\begin{align}
&\rE_{\pos}
\left[\tilde{P}_{\Lambda^{\pos},Z,k,c}(k_k, k_c)\right]
\nonumber \\
=&
\tilde{P}_{\Lambda,Z}(k_k+ k_c)
P_{hg}(k_c|n,l,k_k+ k_c),
\label{1-6-7}
\end{align}
where we used the notation given in (\ref{6-15}).

Now, we consider the case where
Alice and Bob choose the variable $\pos$
and obtain the difference $z_c$ 
between their check bit with the $\times$ basis.
When $\underline{k}\le |z_c| \le \overline{k}$,
the average of Eve's final information is evaluated as
\begin{align}
& \rE_{Y_+}
\left[I([z]\in C_1/C_2(Y_+,nh(\frac{|z_c|}{l}+\delta_{|z_c|})),
\rho_{(\Lambda_t)^{\pos,z},E}^{C_1/C_2(Y)}([z]) )\right]
\nonumber \\
\le &
\overline{h}
\Bigl(
\sum_{k=0}^{n}
\tilde{P}_{(\Lambda_t)^{\pos},Z,k|Z,c}(k |z_c)
f(k,|z_c|~|n,l,\delta_k)
\Bigr)\nonumber \\
&\quad +
n(R-h(|z_c|/l+\delta_k))\nonumber \\
&\quad \cdot \sum_{k=0}^{n}
\tilde{P}_{(\Lambda_t)^{\pos},Z,k|Z,c}(k |z_c)
f(k,|z_c|~|n,l,\delta_k).\label{1-6-1}
\end{align}
When $|z_c| < \underline{k}$,
we obtain
\begin{align}
& \rE_{Y_+}
\left[I([z]\in C_1/C_2(Y_+,
nh(\frac{\underline{k}}{l}+\delta_{\underline{k}})),
\rho_{(\Lambda_t)^{\pos,z},E}^{C_1/C_2(Y)}([z]) )\right]
\nonumber \\
\le &
\overline{h}
\Bigl(
\sum_{k=0}^{n}
\tilde{P}_{(\Lambda_t)^{\pos},Z,k|Z,c}(k |z_c)
f(k,\underline{k}|n,l,\delta_{\underline{k}})
\Bigr)\nonumber \\
&\quad +
n(R-h(\underline{k}/l+\delta_{\underline{k}}))
\nonumber \\
&\hspace{5ex}\cdot \sum_{k=0}^{n}
\tilde{P}_{(\Lambda_t)^{\pos},Z,k|Z,c}(k |z_c)
f(k,\underline{k}|n,l,\delta_{\underline{k}}).\label{1-6-2}
\end{align}
Of course, when $|z_c| > \underline{k}$,
the average of Eve's final information is equal to zero
because any information is discarded in this case.
The inequalities (\ref{1-6-1}) and (\ref{1-6-2})
will be shown in Appendix \ref{3-27-1} 
by using Theorem \ref{th-2}.

Finally, we take the expectation concerning
$z_c$ and $\pos$:
\begin{widetext}
\begin{align}
& \rE_{\pos}\rE_{z_c}
\rE_{Y_+}
\left[I([z]\in C_1/C_2(Y_+,nh(|z_c|/l+\delta_{|z_c|})
),
\rho_{(\Lambda_t)^{\pos,z},E}^{C_1/C_2(Y)}([z]) )\right]
\nonumber \\
\le &
\overline{h}
\Bigl(
\max_j 
\Bigl[
\sum_{k_c=0}^{\underline{k}}
P_{hg}(k_c|n,l,j)
f(j-\underline{k},\underline{k}|n,l,\delta_{\underline{k}})
+
\sum_{k_c=\underline{k}+1}^{\overline{k}}
P_{hg}(k_c|n,l,j)
f(j-k_c,k_c|n,l,\delta_{k_c})
\Bigr]\Bigr)\nonumber\\
&+
\max_j 
\Bigl[
\sum_{k_c=0}^{\underline{k}}
P_{hg}(k_c|n,l,j)
n(R-h(\underline{k}/l+\delta_{\underline{k}}))
f(j-\underline{k},\underline{k}|n,l,\delta_{\underline{k}})
\nonumber \\
&\hspace{10ex} + 
\sum_{k_c=\underline{k}+1}^{\overline{k}}
P_{hg}(k_c|n,l,j)
n(R-h(k_c/l+\delta_{k_c}))
f(j-k_c,k_c|n,l,\delta_{k+c})
\Bigr].\label{1-6-23}
\end{align}
This inequality will be proved in Appendix \ref{a1}.
Hence, we obtain Theorem \ref{th-1}.
Similarly, we have
\begin{align}
& \rE_{\pos}\rE_{z_c}
\rE_{Y_+}
\left[\frac{I([z]\in C_1/C_2(Y_+,nh(|z_c|/l+\delta_{|z_c|})),
\rho_{(\Lambda_t)^{\pos,z},E}^{C_1/C_2(Y)}([z]) )}
{n(R-h(k_\times/l_\times +\delta_{k_\times}))}\right]
\nonumber \\
\le &
\frac{1}{n(R-h(\overline{k}/l_\times +\delta_{\overline{k}}))}
\overline{h}
\Bigl(
\max_j 
\Bigl[
\sum_{k_c=0}^{\underline{k}}
P_{hg}(k_c|n,l,j)
f(k_k,\underline{k}|n,l,\delta_{\underline{k}})
+
\sum_{k_c=\underline{k}+1}^{\overline{k}}
P_{hg}(k_c|n,l,j)
f(k_k,k_c|n,l,\delta_{k_c})
\Bigr]\Bigr)\nonumber\\
&+
\max_j 
\Bigl[
\sum_{k_c=0}^{\underline{k}}
P_{hg}(k_c|n,l,j)
f(k_k,\underline{k}|n,l,\delta_{\underline{k}})
+\sum_{k_c=\underline{k}+1}^{\overline{k}}
P_{hg}(k_c|n,l,j)
f(k_k,k_c|n,l,\delta_{k+c})
\Bigr],
\label{1-6-23-1}
\end{align}
\end{widetext}
This inequality will be proved in Appendix \ref{a1}.
Hence, we obtain Theorem \ref{th-5}.

\section{Security and Phase error}\label{s7}
In this section, we treat the relation between
Eve's information and the phase error.
This relation is one of essential parts 
for Theorem \ref{th-2}.
The purpose of this section is proving the following lemmas\footnote{A similar lemma has been obtained independently by T. Miyadera and Hideki Imai
``On Information-Disturbance Trade-off Theorem,''
{\em Proceedings of ERATO conference on Quantum Information Science 2005,
August 26-30, 2005, Tokyo, Japan, 165 - 166}.}.

\begin{lem}\label{le-1}
Let $\Lambda$ be a generalized Pauli channel on the system
$(\complex^2)^{\otimes n}$.
Then, 
we have
\begin{align}
I(x \in \bF_2^n, \Lambda_E(|x\rangle \langle x|))
\le
\eta_n(1- P_{\Lambda,Z}(0)).
\label{1-5-1}
\end{align}
\end{lem}

Since $1- P_{\Lambda,Z}(0)$ can be regarded as the phase error,
this lemma gives a relation between 
the phase error and Eve's information.

\begin{proof}
The Stinespring representation of $\Lambda$
is given as
$((\complex^2)^{\otimes 2n},U,|\phi\rangle)$:
\begin{align*}
|\phi\rangle&:= \sum_{x,z\in \bF_2^n}\sqrt{P_{\Lambda}(x,z)}|x,z\rangle \\
U&:= \sum_{x,z\in \bF_2^n}
\bX^x \bZ^z \otimes |x,z\rangle \langle x,z|.
\end{align*}
Since
\begin{align*}
& U |x' \rangle \otimes |\phi\rangle=
\sum_{x,z\in \bF_2^n}\sqrt{P_{\Lambda}(x,z)}
(-1)^{x'\cdot z}|x' -x\rangle \otimes |x,z\rangle \\
= &
\sum_{x\in \bF_2^n}
|x' -x\rangle\otimes |\phi_{x,x'}\rangle
\otimes \sqrt{P_{\Lambda,X}(x)}|x\rangle,
\end{align*}
Eve's state can be written as
\begin{align*}
\Lambda_E(|x'\rangle \langle x'|)
= \sum_{x\in \bF_2^n}
P_{\Lambda,X}(x)
|\phi_{x,x'}\rangle\langle \phi_{x,x'}|
\otimes 
|x\rangle \langle x|,
\end{align*}
where
$|\phi_{x,x'}\rangle:=
\sum_{z\in \bF_2^n}\sqrt{P_{\Lambda,Z|X}(z|x)}
(-1)^{x'\cdot z}|z\rangle$.
Since $x'$ obeys the uniform distribution,
\begin{align*}
&I(x \in \bF_2^n, \Lambda_E(|x\rangle \langle x|)) \\
= &
\sum_{x\in \bF_2^n}
P_{\Lambda,X}(x)
H(
\frac{1}{2^n}
\sum_{x'\in \bF_2^n}
|\phi_{x,x'}\rangle\langle \phi_{x,x'}|) \\
=&
\sum_{x\in \bF_2^n}
P_{\Lambda,X}(x)
H(P_{\Lambda,Z|X}(\cdot|x))
\le
H(P_{\Lambda,Z}).
\end{align*}
Hence, using Lemma \ref{le-7},
we obtain (\ref{1-5-1}).
\end{proof}

\begin{lem}\label{le-7}
Let $P=\{P(i)\}$ be a distribution on $\{0, \ldots, d-1\}$.
Then, 
$H(P)\le h(1-P(0))+ \log (d-1)(1-P(0))$.
\end{lem}
\begin{proof}
\begin{align*}
H(P)
=&
-P(0)\log P(0)
-(1-P(0))\log (1-P(0))\\
&-(1-P(0))
\sum_{i=1}^{d-1}
\frac{P(i)}{(1-P(0))}
\log \frac{P(i)}{(1-P(0))}\\
\le &
h(1-P(0))+ \log (d-1)(1-P(0)).
\end{align*}
\end{proof}

\section{Security of known channel}\label{s9}
In this section, we 
treat the security when the channel is known, {\it i.e.},
prove Theorem \ref{th-2} using 
Lemmas \ref{le-1} and \ref{le-2}.
To prove it, for any code $C \subset \bF_2^n$
and any elements $[z]\in \bF_2^n/C^{\perp}$ 
and $[x] \in \bF_2^n/C$, we define
\begin{align}
|x,z\rangle_C := 
\frac{1}{\sqrt{|C|}}\sum_{x'\in C}(-1)^{z \cdot x'} |x+x'\rangle.
\end{align}
Note that this definition does not depend on the choice of
the coset representative elements $z$ ($x$) of $[z]$ ($[x]$).
When we choose $\bF_2^n$ as $C$, the above is the discrete Fourier transform.
Then, we have the following lemma.

\begin{lem}
When two codes $C_1$ and $C_2$ satisfy $C_2 \subset C_1$,
any elements $x\in \bF_2^n$, $[z_1]
\in C_2^{\perp}/C_1^{\perp}$,
and $[z_2]\in \bF_2^n/C_2^{\perp}$ satisfy
\begin{align}
&|x,z_1+z_2\rangle_{C_1}\nonumber \\
=& \frac{1}{\sqrt{|C_1/C_2|}}
\sum_{[x_1]\in C_1/C_2}
(-1)^{(z_1 +z_2)\cdot x_1}
|x+x_1,z_2\rangle_{C_2}.\label{12-30-1}
\end{align}
Note that the RHS does not depend of the choice of
the coset representative elements $x_1$ of $[x_1]$.
\end{lem}
\begin{proof}
\begin{align*}
& \frac{1}{\sqrt{|C_1/C_2|}}
\sum_{[x_1]\in C_1/C_2}
(-1)^{(z_1 +z_2)\cdot x_1}|x+x_1,z_2\rangle_{C_2} \\
=&
\frac{1}{\sqrt{|C_1/C_2|}}
\sum_{[x_1]\in C_1/C_2}
\frac{1}{\sqrt{|C_2|}}
\\
&\cdot \sum_{x_2 \in C_2} 
\quad (-1)^{(z_1 +z_2)\cdot x_1 + z_2 \cdot x_2 }
|x+x_1+x_2\rangle.
\end{align*}
Since 
$(z_1 +z_2 )\cdot (x_1 + x_2 )=
(z_1 +z_2)\cdot x_1 + z_2 \cdot x_2 $,
we obtain (\ref{12-30-1}).
\end{proof}
\begin{lem}\label{le-3}
\begin{align}
\sum_{x_1 \in C_1}
|x+x_1 \rangle \langle x+x_1|
=
\sum_{[z_1] \in \bF_2^n/C_1^{\perp}}
|x,z_1 \rangle_{C_1}~_{C_1} \langle x,z_1|.
\end{align}
\end{lem}
\begin{proof}
From the definition of $|x,z_1 \rangle_{C_1}$,
we have
\begin{align*}
&\sum_{[z_1] \in \bF_2^n/C_1^{\perp}}
|x,z_1 \rangle_{C_1}~_{C_1} \langle x,z_1| \\
= &
\frac{1}{|C_1|}
\sum_{[z_1] \in \bF_2^n/C_1^{\perp}}
\sum_{x'\in C_1}
\sum_{x''\in C_1}
(-1)^{z_1 \cdot (x'+ x'')} 
|x+x''\rangle\langle x+x'|\\
=& 
\frac{1}{|C_1|}
\sum_{[z_1] \in \bF_2^n/C_1^{\perp}}
\sum_{x'\in C_1}
\sum_{x''\in C_1} \\
& \hspace{5ex}(-1)^{z_1 \cdot (x'+x' + (x''- x'))} 
|x+x'+x''-x'\rangle\langle x+x'|\\
= &
\frac{1}{|C_1|}
\sum_{[z_1] \in \bF_2^n/C_1^{\perp}}
\sum_{x'\in C_1}
\sum_{y\in C_1}
(-1)^{z_1 \cdot y} 
|x+x'+y\rangle\langle x+x'| \\
=&
\sum_{x'\in C_1}
|x+x'\rangle\langle x+x'|,
\end{align*}
because $y\in C_1$ satisfies
\begin{align*}
\frac{1}{|C_1|}
\sum_{[z_1] \in \bF_2^n/C_1^{\perp}}
(-1)^{z_1 \cdot y} = 
\left\{
\begin{array}{ll}
1 & \hbox{ if } y =0 \\
0 & \hbox{ if } y \neq 0 .
\end{array}
\right.
\end{align*}
\end{proof}

Now, we define
the minimum error:
\begin{align*}
&P([z_1]\in  C_2^{\perp}/C_1^{\perp},
\Lambda(|0, z_1+ z_2\rangle_{C_1}~ _{C_1} \langle 0, z_1+ z_2|))\\
:=&\min_{M}
\Bigl(1- \\
& 
\sum_{[z_1]\in  C_2^{\perp}/C_1^{\perp}}
\frac{
\Tr M_{[z_1]}
\Lambda(|0, z_1+ z_2\rangle_{C_1}~ _{C_1} \langle 0, z_1+ z_2|))
}
{|C_2^{\perp}/C_1^{\perp}|}
\Bigr),
\end{align*}
where 
$M$ is a POVM $\{M_{[z_1]}\}_{[z_1]\in  C_2^{\perp}/C_1^{\perp}}$.
Then, we have the following evaluation.
\begin{lem}\label{l1-6-16}
\begin{align}
& I([x_1]\in C_1/C_2, 
\Lambda_E(
\frac{1}{|C_2|}\sum_{x_2 \in C_2}
|x_1+x_2\rangle \langle x_1+x_2|))
\nonumber \\
\le &
\eta_{m-s}
\sum_{[z_2] \in \bF_2^n/C_2^{\perp}}
\frac{1}{|C_2|}\nonumber \\
&\cdot
P(z_1\in  C_2^{\perp}/C_1^{\perp},
\Lambda(|0, z_1+ z_2\rangle_{C_1}~ _{C_1} \langle 0, z_1+ z_2|)),
\label{1-6-16}
\end{align}
where
$m=\dim C_1$ and $s=\dim C_2$.
\end{lem}
\begin{proof}
Using Lemma \ref{le-3} and 
the convexity of mutual information,
we have
\begin{align}
& I([x_1]\in C_1/C_2, 
\Lambda_E(
\frac{1}{|C_2|}\sum_{x_2 \in C_2}
|x_1+x_2\rangle \langle x_1+x_2|))
\nonumber \\
=&
I([x_1]\in C_1/C_2, 
\Lambda_E(
\frac{1}{|C_2|}\sum_{[z_2] \in \bF_2^n/C_2^{\perp}}
|x_1,z_2\rangle_{C_2}~ _{C_2} \langle x_1,z_2|))\nonumber \\
\le &
\frac{1}{|C_2|}\sum_{[z_2] \in \bF_2^n/C_2^{\perp}}
I([x_1]\in C_1/C_2, 
\Lambda_E(
|x_1,z_2\rangle_{C_2}~ _{C_2} \langle x_1,z_2|)). \label{1-6-20}
\end{align}
Applying Lemma \ref{le-1},
we have
\begin{align}
& 
I(x_1\in C_1/C_2, 
(\Lambda_E
(|x_1,z_2\rangle_{C_2}~ _{C_2}\langle x_1,z_2|))\nonumber\\
\le &
\eta_{m-s}
P(z_1\in  C_2^{\perp}/C_1^{\perp},
\Lambda(|0, z_1+ z_2\rangle_{C_1}~ _{C_1} \langle 0, z_1+ z_2|)).
\nonumber
\end{align}
From (\ref{1-6-20}), the concavity of $\eta_{m-s}$ implies
\begin{align*}
& I([x_1]\in C_1/C_2, 
\Lambda_E(
\frac{1}{|C_2|}\sum_{x_2 \in C_2}
|x_1+x_2\rangle \langle x_1+x_2|))\\
\le &
\frac{1}{|C_2|}
\sum_{[z_2] \in \bF_2^n/C_2^{\perp}}
\eta_{m-s}
P(z_1\in  C_2^{\perp}/C_1^{\perp},
\Lambda(|0, z_1+ z_2\rangle_{C_1}~ _{C_1} \langle 0, z_1+ z_2|))\\
\le &
\eta_{m-s}
\frac{1}{|C_2|}
\sum_{[z_2] \in \bF_2^n/C_2^{\perp}}
P(z_1\in  C_2^{\perp}/C_1^{\perp},
\Lambda(|0, z_1+ z_2\rangle_{C_1}~ _{C_1} \langle 0, z_1+ z_2|)).
\end{align*}
\end{proof}

Since $\Lambda$ is a generalized Pauli channel,
any coset $[x_0] \in \bF_2^n/C_1$ satisfies
$\Lambda(|x_0, z_1+ z_2\rangle_{C_1}~ _{C_1} \langle x_0, z_1+ z_2|))
=
\bX^{x_0}
\Lambda(|0, z_1+ z_2\rangle_{C_1}~ _{C_1} \langle 0, z_1+ z_2|))
(\bX^{x_0})^\dagger$.
Hence, 
\begin{align*}
& 
P([z_1]\in  C_2^{\perp}/C_1^{\perp},
\Lambda(|0, z_1+ z_2\rangle_{C_1}~ _{C_1} \langle 0, z_1+ z_2|))\\
=&
P([z_1]\in  C_2^{\perp}/C_1^{\perp},
\Lambda(|x_0, z_1+ z_2\rangle_{C_1}~ _{C_1} \langle x_0, z_1+ z_2|)).
\end{align*}
Thus, 
\begin{align}
& 
P(z_1\in  C_2^{\perp}/C_1^{\perp},
\Lambda(|0, z_1+ z_2\rangle_{C_1}~ _{C_1} \langle 0, z_1+ z_2|))\nonumber\\
=&
\frac{|C_1|}{2^n}\sum_{[x_0] \in \bF_2^n/C_1}\nonumber\\
& P(z_1\in  C_2^{\perp}/C_1^{\perp},
\Lambda(|x_0, z_1+ z_2\rangle_{C_1}~ _{C_1} \langle x_0, z_1+ z_2|))\nonumber\\
\le &
P\Bigl(z_1\in  C_2^{\perp}/C_1^{\perp},\nonumber\\
&\frac{|C_1|}{2^n}\sum_{[x_0] \in \bF_2^n/C_1}
\Lambda(|x_0, z_1+ z_2\rangle_{C_1}~ _{C_1} \langle x_0, z_1+ z_2|)
\Bigr)\nonumber\\
=&
P\Bigl(z_1\in  C_2^{\perp}/C_1^{\perp},\nonumber\\
&\Lambda(
\frac{|C_1|}{2^n}\sum_{z_0 \in C_1^{\perp}}
|z_0 + z_1+ z_2\rangle_{\bF_2^n} ~_{\bF_2^n}\langle z_0+ z_1+ z_2|)
\Bigr).
\label{1-6-14}
\end{align}
Now, we focus on 
the step (ix) and the subcode $G(C_1) C_2(Y,s)\subset C_1$,
and abbreviate $G(C_1) C_2(Y,s)$ to $C_2(Y,s)$.
Then, the dual code $C_2(Y,s)^{\perp}$
satisfies $C_1^{\perp}\subset C_2(Y,s)^{\perp}$
and the condition of $C_2(X)$ in Lemma \ref{le-2}
when 
$t$, $l$ and $C_1$ in Lemma \ref{le-2}
is given by 
$n-v$, $v-s$, and $C_1^{\perp}$, respectively.
Then, $n-(l+t)$ in Lemma \ref{le-2} is given by $s$.
Since the generalized Pauli channel can be regarded
as the additive channel,
we can apply Lemma \ref{le-2}.
Hence,
\begin{align}
&\rE_{Y}
\Bigl[\frac{1}{|C_2(Y,s)|}\sum_{[z_2] \in \bF_2^n/C_2(Y,s)^{\perp}}\nonumber\\
& P\Bigl(z_1\in  C_2(Y,s)^{\perp}/C_1^{\perp},\nonumber\\
&\quad \Lambda_E(
\frac{|C_1|}{2^n}
\sum_{z_0 \in C_1^{\perp}}
|z_0 + z_1+ z_2\rangle_{\bF_2^n} ~_{\bF_2^n}\langle z_0+ z_1+ z_2|)
\Bigr)\Bigr]\nonumber\\
\le &
\sum_{k=0}^{n}
\tilde{P}_{W_{t}}(k)
g(2^{-s}|n,k).
\label{1-6-15}
\end{align}

From (\ref{1-6-14}), (\ref{1-6-15}), and (\ref{1-6-16}),
the convexity of $\eta_{m-s}$ yields that
\begin{align*}
& \rE_{Y}
\Bigl[\frac{1}{|C_2|}\sum_{[z_2] \in \bF_2^n/C_2^{\perp}} \\
& \hspace{5ex}I([x_1]\in C_1/C_2, 
\Lambda_E(
|x_1,z_2\rangle_{C_2}~ _{C_2} \langle x_1,z_2|) )\Bigr]\\
\le &
\eta_{m-s }
\Bigl(\rE_{Y}\Bigl[
\frac{1}{|C_2|}\sum_{[z_2] \in \bF_2^n/C_2^{\perp}}\\
&\hspace{5ex}
P\bigl(z_1\in  C_2^{\perp}/C_1^{\perp}, 
\Lambda(|0, z_1+ z_2\rangle_{C_1}~ _{C_1} \langle 0, z_1+ z_2|\bigr)
\Bigr]\Bigr)
\\
\le &
\eta_{m-s }\Bigl(
\sum_{k=0}^{\lfloor n/2 \rfloor}
\tilde{P}_{W_{t}}(k)
g(2^{-s}|n,k)
\Bigr).
\end{align*}
Therefore, from (\ref{1-6-20}), we obtain Theorem \ref{th-2}.

\section{Optimal Attack}\label{s10}
In this section, we prove
that there exists a collective attack attaining the 
the exponential rate (\ref{1-29-2}) under a condition.
Indeed, it is not so easy to evaluate 
$\max I(\overline{Z},Z_{E})$.
Hence, we treat $I_H(\overline{Z},Z_{E})$ 
instead of $I(\overline{Z},Z_{E})$.

\begin{lem}\label{2-5-5}
Assume that 
the sequence of codes $C_{1,n,k_+}$ satisfies 
\begin{align}
\max_{k \le (\overline{p}+\epsilon(\overline{p}))
n} P_{e,W_{k,n}} (C_{1,n}^\perp) \to 0 \label{2-5-4}
\end{align}
where the channel $W_{k,n}$ is defined on $\bF_2^n$ as
$P_{W_{k,n}}(j)= \delta_{k,j}$.
Then, we have
\begin{align}
&\lim \frac{-r}{n}\log 
\rE_{k_\times|\pos_\times,Y_ +,\pos_+,k_+,Y_\times} 
\left[\max_{{\cal E}}
I_H(\overline{Z},Z_{E}) \right]\nonumber \\
\le &
\min_{p \in [\underline{p},\overline{p}]}
h(p + r \epsilon(p))-(1-r)h(p)-r h(p + \epsilon(p)),
\label{2-5-1}
\end{align}
where the maximum is taken concerning Eve's operation ${\cal E}$.
Note that, the above inequality holds for
any fixed variable $Y_+$.
\end{lem}
Hence, if $\epsilon(p)$ is sufficiently small
and the sequence of codes $C_{1,n}$ satisfies
the condition (\ref{2-5-4}),
we have
\begin{align}
&\lim \frac{-r}{n}\log 
\rE_{\pos_\times,k_\times,Y_ +|\pos_+,k_+,Y_\times} 
\left[\max 
I_H(\overline{Z},Z_{E}) \right]\nonumber \\
=& h(p + r \epsilon(p))-(1-r)h(p)-r h(p + \epsilon(p)).
\end{align}
This indicates that the method of randomly choosing code $C_2$
is optimal in the sense of large deviation.

In this lemma, we assume the condition (\ref{2-5-4}).
Indeed, we need some conditions in Lemma \ref{2-5-5}.
For example, consider the code 
$C_1$ that consists of the elements $x$ whose the first $n-m$
components is zero.
In this case, the following proof is not valid.
Indeed, when the limit 
$\lim_{n \to \infty}\frac{1}{n}\log |C_{1,n}|$
is greater than $h(\overline{p} +\epsilon(\overline{p}))$
and we choose $C_{1,n}$ randomly,
the condition (\ref{2-5-4}) holds.
Hence, the condition (\ref{2-5-4}) is not so unnatural.
However, a more natural condition is needed.

As is shown later, 
the exponential rate 
$\min_{p \in [\underline{p},\overline{p}]}
h(p + r \epsilon(p))-(1-r)h(p)-r h(p + \epsilon(p))$
can be attained by a collective attack,
in which 
Eve's is allowed only individual unitary operations
to quantum states sent by Alice and 
any global generalized measurement on the Eve's 
local states.
Hence, 
the exponential rate of Eve's information
cannot be improved by
any collective attack, in which 
Eve's is allowed to use any unitary operation
to all quantum states sent by Alice.

Now, we construct Eve's strategy attaining 
the bound
$\min_{p \in [\underline{p},\overline{p}]}
h(p + r \epsilon(p))-(1-r)h(p)-r h(p + \epsilon(p))$
and prove (\ref{2-5-1}).
Choose $p_0:=\argmin_{p \in [\underline{p},\overline{p}]}
h(p + r \epsilon(p))-(1-r)h(p)-r h(p + \epsilon(p))$.
Eve performs a unitary action $U_{p_0+r \epsilon(p_0)}
$
\begin{align*}
U_p |x\rangle \otimes |0 \rangle_E:= 
\sqrt{p} |x \rangle \otimes |0 \rangle_E
+
(-1)^{x}
\sqrt{1-p} |x \rangle \otimes |1 \rangle_E
\end{align*}
for a every qubit,
where $|x \rangle_E$ is Eve's state.

We define the unitary $\tilde{U}_k$:
\begin{align*}
\tilde{U}_k^n |x\rangle \otimes |0 \rangle_E:= 
\sqrt{
\frac{1}{
\genfrac{(}{)}{0pt}{}{n}{k}
}}
\sum_{
y\in \bF_2^n:|x|=k}
(-1)^{x\cdot y}
 |x \rangle \otimes |y \rangle_E.
\end{align*}
We can easily show that
\begin{align*}
& H\Bigl(
\Lambda_{E,k}^k\Bigl(
\sum_{x\in C_{1,n}}|x \rangle\langle x|\Bigr)
\Bigr)\nonumber \\
=&
H\Bigl(
\Lambda_{E,k}^k\Bigl(
\sum_{x\in C_{1,n}}|y+x \rangle\langle y+x|\Bigr)
\Bigr) \hbox{ for } y \in \bF_2^n.
\end{align*}
Hence, applying Lemma \ref{l1-6-16}
to the case of $C_1=\bF_2^n, C_2=C_{1,n}$,
we have
\begin{align*}
& \log \genfrac{(}{)}{0pt}{}{n}{k}
-
H\Bigl(
\Lambda_{E,k}^n\Bigl(
\sum_{x\in C_{1,n}}|x \rangle\langle x|\Bigr)
\Bigr) \\
= &
H\Bigl(
\Lambda_{E,k}^n\Bigl(
\sum_{x\in \bF_2^n}|x \rangle\langle x|\Bigr)
\Bigr)  -
H\Bigl(
\Lambda_{E,k}^n\Bigl(
\sum_{x\in C_{1,n}}|x \rangle\langle x|\Bigr)
\Bigr) \\
\le &
\overline{h}
(P_{e,W_{k,n}} (C_{1,n}^\perp) )
+
\log |C_{1,n}|
P_{e,W_{k,n}} (C_{1,n}^\perp) ,
\end{align*}
where
\begin{align*}
\Lambda_{E,k}^n(\rho):=
\Tr_B \tilde{U}_k^n 
(\rho \otimes |0 \rangle_E~_E\langle 0|)
(\tilde{U}_k^n )^{\dagger}.
\end{align*}

Now, we evaluate Eve's information.
In this case, the subcode $C_{2,n,k_\times}$ depends on 
the outcome $k_\times$.
Taking the pinching map:
$\rho\mapsto
\sum_k
P_{n,k}
\rho
P_{n,k}$ ($
P_{n,k}$ is the projection to the space spanned by
$\{|x\rangle\}_{|x|=k}$),
we have
\begin{widetext}
\begin{align*}
& H\Bigl(\rE_{k_{\times}}
\Bigl[\Lambda_E^{\otimes n}
\Bigl(
\sum_{x\in C_{1,n}}|x \rangle\langle x|
\Bigr)\Bigr] \Bigr)
-
\sum_{[x]_2\in C_{1,n}/C_{2,n,k_\times}}
H\Bigl(\Lambda_E^{\otimes n}
\Bigl(
\sum_{y\in C_{2,n,k_\times}}|x+y \rangle\langle x+y|
\Bigr)\Bigr)
\\
\ge &
H\Bigl(\rE_{k_{\times},k}
\Bigl[\Lambda_{E,k}^n\Bigl(
\sum_{x\in C_{1,n}}|x \rangle\langle x|
\Bigr)\Bigr] \Bigr)
-
\sum_{[x]_2\in C_{1,n}/C_{2,n,k_\times}}
H\Bigl(\Lambda_{E,k}^n\Bigl(
\sum_{y\in C_{2,n,k_\times}}|x+y \rangle\langle x+y|
\Bigr)\Bigr),
\end{align*}
where $k$ is the random variable
with distribution 
$\tilde{P}(k):=
\genfrac{(}{)}{0pt}{}{n}{k}
(p_0+r \epsilon(p_0))^k
(1-p_0-r \epsilon(p_0))^{n-k}$.

When $k=n(p_0+ \epsilon(p_0)+\epsilon)$, $k_\times= p_0$,
we have
\begin{align*}
& \frac{1}{n}
H\Bigl(\Lambda_{E,k}^n\Bigl(
\sum_{x\in C_{1,n}}|x \rangle\langle x|\Bigr)\Bigr)
-
\frac{1}{n}
\sum_{[x]_2\in C_{1,n}/C_{2,n,k_\times}}
H\Bigl(\Lambda_{E,k}^n\Bigl(
\sum_{y\in C_{2,n,k_\times}}|x+y \rangle\langle x+y|
\Bigr)\Bigr)\\
\ge &
\frac{1}{n}
\left(
\log \genfrac{(}{)}{0pt}{}{n}{k}
-
\overline{h}
(P_{e,W_{k,n}} (C_{1,n}^\perp) )
-
\log |C_{1,n}|
P_{e,W_{k,n}} (C_{1,n}^\perp) 
- \log |C_{2,n,k_\times}|
\right)\\
= &
\frac{1}{n}
\left(
\log \genfrac{(}{)}{0pt}{}{n}{k}
- n h(\frac{k_\times}{n}+\epsilon(\frac{k_\times}{n}))
-
\overline{h}
(P_{e,W_{k,n}} (C_{1,n}^\perp) )
-
\log |C_{1,n}|
P_{e,W_{k,n}} (C_{1,n}^\perp) 
\right)\\
\to &
h(p_0+ \epsilon(p_0)+\epsilon)-h(p_0+ \epsilon(p_0))
\quad \hbox{as} \quad n \to \infty.
\end{align*}
Hence, Eve's information can be bounded as
\begin{align*}
& \rE_{k_{\times}}
\left[H(\Lambda_E^{\otimes n}
(
\sum_{x\in C_{1,n}}|x \rangle\langle x|)
))
-
\sum_{[x]_2\in C_{1,n}/C_{2,n,k_\times}}
H(\Lambda_E^{\otimes n}
(
\sum_{y\in C_{2,n,k_\times}}|x+y \rangle\langle x+y|) )\right]\\
\ge &
\genfrac{(}{)}{0pt}{}{n}{n(p_0+ \epsilon(p_0)+\epsilon)}
(p_0+r \epsilon(p_0))^{n(p_0+ \epsilon(p_0)+\epsilon)}
(1-p_0-r \epsilon(p_0))^{n-n(p_0+ \epsilon(p_0)+\epsilon)} \\
& \cdot \genfrac{(}{)}{0pt}{}{l}{l p_0}
(p_0+r \epsilon(p_0))^{l p_0}
(1-p_0-r \epsilon(p_0))^{l(1-p_0)}
(n (h(p_0+ \epsilon(p_0)+\epsilon)-h(p_0+ \epsilon(p_0)))+o(n)) \\
\ge &
\frac{n (h(p_0+ \epsilon(p_0)+\epsilon)-h(p_0+ \epsilon(p_0)))+o(n)}
{(n+1)^2}
2^{-n d(p_0+ \epsilon(p_0)+\epsilon\|p_0+ r \epsilon(p_0))
-l d(p_0 \|p_0+ r \epsilon(p_0))}.
\end{align*}
\end{widetext}
Thus, we obtain
\begin{align*}
& \lim \frac{-r}{n}\log 
\rE_{k_\times|\pos_\times,Y_ +,\pos_+,k_+,Y_\times} 
\left[\max I_H\left( \overline{Z}_+ ,Z_E \right) \right] \\
\le &
h(p_0 + r \epsilon(p_0))-(1-r)h(p_0)-r h(p_0 + \epsilon(p_0)
+ \epsilon).
\end{align*}
Taking the limit $\epsilon \to 0$, we
obtain (\ref{2-5-1}).

\section{Conclusion}
In this paper, we obtained a practical evaluation of security of 
quantum key distribution.
This bound improves existing bounds.
In order to guarantee the security of 
implemented QKD system,
we need a tighter bound in the finite-coding length.
Hence, our bound is useful for 
guaranteeing the security of 
quantum key distribution with perfect single photon source.
However, for a precise evaluation, 
we have to treat hypergeometric distributions,
because our bound contains hypergeometric distributions.
Hence, it is needed to calculate these bounds 
by numerical analysis based on several calculations of
hypergeometric distributions.

We also derived the exponential rate of our bound as (\ref{1-29-2}),
and proved its optimality with in the sense of Holevo information
with a class of one-way communication
when $C_p$ is less than the critical case.
However, our condition for our code is not sufficiently natural.
Hence, it is required to prove this optimality 
under a more natural condition. 
One candidate of a more natural condition is 
\begin{align}
\max_{k \le 
\bar{h}^{-1}
(1-h(\overline{p}+\epsilon(\overline{p})))n} 
P_{e,W_{k,n}} (C_{1,n}^\perp) \to 0 \label{2-5-4-1}.
\end{align}
Hence, it is a future problem to show the optimality 
under the above condition.

Further, we assumed that perfect single photon source.
One idea for the weak coherent case
is the decoy method \cite{hwang},
which is based on 
the observation of the security with imperfect devices\cite{GLLP}.
However, any existing paper \cite{lo4,wang2d,lolo}
of the decoy method does not discuss the degree of Eve's information 
in the framework of finite coding-length, precisely.
Hence, it is required to extend our result to the weak coherent case
with the decoy method.

\section*{Acknowledgments}
The author would like to thank Professor Hiroshi Imai of the
ERATO-SORST, QCI project for support.
He is grateful to Professor Hiroshi Imai, Dr. Akihisa Tomita, 
Professor Keiji Matsumoto,
and Mr. Jun Hasewaga for useful discussions.
He is also benefited by referee for pointing out several mistakes
in the first version.

\appendix

\section{Derivation of (\ref{1-29-1})}\label{a2}
Now, we prove (\ref{1-29-1}).
In the following, we denote $n+l$ by $m$ and fix 
$p \in [\underline{p}, \overline{p}]$.
We treat the case of $j = p m$, and 
define the number $k_p(m):= 
\max \{ k| h(\frac{pm -k}{n})-h (\frac{k}{l}+\delta_k)
\ge 0\}$.
In this case, 
the first term of 
$\tilde{P}(\delta,n,l,\underline{k},\overline{k})$
goes to $0$.
Hence, we focus on 
the second term of 
$\tilde{P}(\delta,n,l,\underline{k},\overline{k})$,
which is divided as 
\begin{align*}
&\tilde{P}_2(\delta,n,l,\underline{k},\overline{k}) \\
:=&
\sum_{k=0}^{\underline{k}}
P_{hg}(k|n,l,j)
f(j-\underline{k},\underline{k}|n,l,\delta_{k_\times}) \\
&+\sum_{k=\underline{k}+1}^{\overline{k}}
P_{hg}(k|n,l,j)
f(j-k,k|n,l,\delta_{k_\times}) \\
=&
\sum_{k=0}^{k_p(m)}
P_{hg}(k|n,l,j) \\
& +
\sum_{k=k_p(m)+1}^{\overline{k}}
P_{hg}(k|n,l,j)
f(j-k,k|n,l,\delta_{k_\times}).
\end{align*}
Since
$h(\frac{pm -k_p(m)}{n})=h (\frac{k_p(m)}{l}+\delta_k)$,
we have
$k_p(m)= p l - \frac{n l}{m} \delta_k$.
Using the relation $\delta_k = \frac{\tilde{\epsilon}_{k/l}}{\sqrt{m}}$
and 
the continuity of $C_p$,
we have $k_p(m)=(1-r)pm-r(1-r)\tilde{\epsilon}(p) \sqrt{m}+o(\sqrt{m})$.
The average of $k$ is $\frac{lj}{n+l}=(1-r)p m$
and the variance of $k$ is
$\frac{jln(n+l-j)}{(n+l)^2 (n+l-1)}=\frac{r(1-r)p(1-p)m}{1-\frac{1}{m}}$.
Hence, 
$\frac{k_p(m)-(1-r)p m}
{\sqrt{\frac{r(1-r)p(1-p)m}{1-\frac{1}{m}}}}
\to-
\frac{\sqrt{r(1-r)}}{\sqrt{p(1-p)}}\tilde{\epsilon}(p)$.
Thus, we have
\begin{align*}
\sum_{k=0}^{k_p(m)}P_{hg}(k|n,l,j)
=
\Phi\left(-
\frac{\sqrt{r(1-r)}}{\sqrt{p(1-p)}}\tilde{\epsilon}(p)
\right).
\end{align*}

When $k \ge k_p(m)$, we can approximate the difference as
\begin{align*}
h(\frac{j-k}{n})-h(\frac{k}{l}+\delta_k)
\cong -h'(p) \frac{l+n}{ln}(k- k_p(m)).
\end{align*}
Hence,
\begin{align*}
&\sum_{k=k_p(m)+1}^{\overline{k}}
P_{hg}(k|n,l,j)
f(j-k,k|n,l,\delta_{k_\times}) \\
\cong &
\frac{1}{\sqrt{2\pi \frac{r(1-r)p(1-p)m}{1-\frac{1}{m}}}} \\
& \cdot \int_{k_p(m)}^{\overline{k}}
e^{-\frac{(x-(1-r)p m)^2}
{2\frac{r(1-r)p(1-p)m}{1-\frac{1}{m}}}}
2^{-n h'(p) \frac{l+n}{ln}(x- k_p(m))}
d x \\
\cong &
\frac{1}{2\pi}
\int_{-
\frac{\sqrt{r(1-r)}}{\sqrt{p(1-p)}}
\tilde{\epsilon}(p)
}
^{+\infty}
e^{-\frac{x^2}{2}} 
2^{
- \sqrt{m}h'(p)
\frac{\sqrt{r p(1-p)}}{\sqrt{1-r}}
(y+
\frac{\sqrt{r(1-r)}}{\sqrt{p(1-p)}}
\tilde{\epsilon}(p)
)
}\\
& \quad \cdot
\sqrt{r(1-r)p(1-p) m}
d y \\
\to & 0 \hbox{ as } m \to \infty,
\end{align*}
where $y=
\frac{x-(1-r)p m}{\sqrt{
r(1-r)p(1-p)m}}$.

Next, we consider the case when $\frac{j}{m}$ is strictly smaller 
than $\underline{p}$.
The value $-h(\frac{j-\underline{k}}{n})+
h(\frac{\underline{k}}{l}+\delta_{\underline{k}})$
is strictly positive
and 
$-h(\frac{j-k}{n})+
h(\frac{k}{l}+\delta_{k})$
is smaller than this value if $k \ge \underline{k}$.
Hence, 
$\tilde{P}_2(\delta,n,l,\underline{k},\overline{k})$
goes to $0$.

Finally, we consider the case when $\frac{j}{m}$ is strictly
greater than 
$\overline{p}$.
In this case,
as is mentioned in \ref{a3},
the probability that $k$ is greater than $\overline{k}$ 
exponentially goes to $0$.
Hence, in this case
$\tilde{P}(\delta,n,l,\underline{k},\overline{k})$
goes to $0$.
Therefore, we obtain (\ref{1-29-1}).

\section{Derivation of (\ref{1-29-2})}\label{a3}
From (\ref{1-16-30}),
we have
\begin{align*}
&\frac{1}{(n+1)(l+1)}
2^{
l h(\frac{k}{l})+ n h(\frac{j-k}{n})
- (n+l)h(\frac{j}{n+l})}
\le
P_{hg}(k|n,l,j) \\
=&\frac{
\genfrac{(}{)}{0pt}{}{l}{k}
\genfrac{(}{)}{0pt}{}{n}{j-k}
}{
\genfrac{(}{)}{0pt}{}{n+l}{j}}
\le
(n+l+1)
2^{
l h(\frac{k}{l})+ n h(\frac{j-k}{n})
- (n+l)h(\frac{j}{n+l})}.
\end{align*}
Hence,
\begin{align*}
& \max_j \sum_{k=0}^{\underline{k}}
\Bigl[P_{hg}(k|n,l,j)
f(j-\underline{k},\underline{k}|n,l,\delta_{k_\times}) \\
&\quad  +
\sum_{k=\underline{k}+1}^{\overline{k}}
P_{hg}(k|n,l,j)
f(j-k,k|n,l,\delta_{k_\times}) \Bigr]\\
\le &
\overline{k}
(n+l+1)\\
&\cdot 2^{
\max_{j,k}
l h(\frac{k}{l})+ n h(\frac{j-k}{n})- (n+l)h(\frac{j}{n+l})
- n [h (\frac{k}{l}+ \delta_k)-h(\frac{j-k}{n})]_+}.
\end{align*}
Further,
\begin{align*}
&\max_j \Bigl[\sum_{k=0}^{\underline{k}}
P_{hg}(k|n,l,j)
f(j-\underline{k},\underline{k}|n,l,\delta_{k_\times})\\
& \hspace{9ex} \cdot n(R-h(\frac{\underline{k}}{l} + \delta_{k_\times}))
\\
&%
+
\sum_{k=\underline{k}+1}^{\overline{k}}
P_{hg}(k|n,l,j)
f(j-k,k|n,l,\delta_{k_\times}) \\
& \hspace{9ex}\cdot n(R-h(\frac{k}{l}+ \delta_{k_\times}))
\Bigr],\\
\le &
n (R-h(\underline{p}+\epsilon(\underline{p})))
\overline{k}
(n+l+1) \\
& \cdot 2^{
\max_{j,k}
l h(\frac{k}{l})+ n h(\frac{j-k}{n})- (n+l)h(\frac{j}{n+l})
- n [h (\frac{k}{l}+ \delta_k)-h(\frac{j-k}{n})]_+}.
\end{align*}
Thus, 
substituting
$p=\frac{k}{l}, r=\frac{n}{n+l}, \epsilon(p)=\delta_k,
\epsilon'= \frac{k}{l}-\delta_k-\frac{j-k}{n}$,
we obtain (\ref{1-29-3}).
Since
\begin{align}
& \frac{-r}{n}
\max_{j,k}
\Bigl[ l h(\frac{k}{l})+ n h(\frac{j-k}{n})- (n+l)h(\frac{j}{n+l}) \nonumber \\
&\hspace{10ex} - n [h (\frac{k}{l}+ \delta_k)-h(\frac{j-k}{n})]_+\Bigr]\nonumber \\
\le &
E(\epsilon,r,\underline{p},\overline{p}),
\label{1-29-5}
\end{align}
we obtain the part $\le$ in (\ref{1-29-2}).

Conversely,
\begin{align*}
& \max_j \Bigl[\sum_{k=0}^{\underline{k}}
P_{hg}(k|n,l,j)
f(j-\underline{k},\underline{k}|n,l,\delta_{k_\times}) \\
&\hspace{10ex} +
\sum_{k=\underline{k}+1}^{\overline{k}}
P_{hg}(k|n,l,j)
f(j-k,k|n,l,\delta_{k_\times}) \Bigr]\\
\ge &
\frac{
2^{
\max_{j,k}
l h(\frac{k}{l})+ n h(\frac{j-k}{n})- (n+l)h(\frac{j}{n+l})
- n [h (\frac{k}{l}+ \delta_k)-h(\frac{j-k}{n})]_+}
}{(n+1)(l+1)}.
\end{align*}
Since 
the equality in (\ref{1-29-5}) holds 
in the limit $n \to \infty$,
we obtain the part $\ge$ in (\ref{1-29-2}).

\section{Proof of (\ref{1-6-1}) and (\ref{1-6-2})}\label{3-27-1}
When Alice sends the classical information
$x+ X_+$ ($x= G(C_1)Z$),
the probability that Bob obtains the local signal
$x_b:= x+ X_+ -\tilde{X}_+$ 
is 
\begin{widetext}
\begin{align}
&
\frac{\displaystyle
\Tr\frac{1}{2^{n+l}}
\sum_{x_k'\in \bF_2^n}\sum_{z_c'\in \bF_2^l}
\Lambda^{\pos}
(|x_k'\rangle \langle x_k'|\otimes 
|z_c'\rangle_{\bF_2^n}~_{\bF_2^n} \langle z_c'|)
|x_k'+ x-x_b\rangle \langle x_k'+x-x_b|\otimes
|z_c'-z\rangle_{\bF_2^n}~_{\bF_2^n} \langle z_c'-z|)
}{
\Tr \frac{1}{2^{l}}
\sum_{x_c'\in \bF_2^l}
\Lambda^{\pos}
( \rho_{\mix,n}\otimes 
|z_c'\rangle_{\bF_2^n}~_{\bF_2^n} \langle z_c'|)
I\otimes|z_c'-z\rangle_{\bF_2^n}~_{\bF_2^n} \langle z_c'-z|)}\nonumber\\
=&
\frac{\displaystyle\Tr \frac{1}{2^{n+l}}
\sum_{x_k''\in \bF_2^n}\sum_{z_c'\in \bF_2^l}
\Lambda^{\pos}
(|x_k''-x_b\rangle \langle x_k''-x_b|\otimes 
|z_c'\rangle_{\bF_2^n}~_{\bF_2^n} \langle z_c'|)
|x_k''-x_b\rangle \langle x_k''-x_b|\otimes
|z_c'-z\rangle_{\bF_2^n}~_{\bF_2^n} \langle z_c'-z|)
}{\Tr
\frac{1}{2^{l}}
\sum_{x_c'\in \bF_2^l}
\Lambda^{\pos}
( \rho_{\mix,n}\otimes 
|z_c'\rangle_{\bF_2^n}~_{\bF_2^n} \langle z_c'|)
I\otimes|z_c'-z\rangle_{\bF_2^n}~_{\bF_2^n} \langle z_c'-z|)}\nonumber\\
=&
\frac{\Tr \frac{1}{2^{n+l}}
\sum_{x_k''\in \bF_2^n}\sum_{z_c'\in \bF_2^l}
(\Lambda^{\pos})^{(x_k''0,0 z_c') }
(|-x_b\rangle \langle -x_b|\otimes 
|0\rangle_{\bF_2^n}~_{\bF_2^n} \langle 0|)
|-x_b\rangle \langle -x_b|\otimes
|-z\rangle_{\bF_2^n}~_{\bF_2^n} \langle -z|)
}{\Tr
\frac{1}{2^{l}}
\sum_{x_c'\in \bF_2^l}
\Lambda^{\pos}
( \rho_{\mix,n}\otimes 
|0\rangle_{\bF_2^n}~_{\bF_2^n} \langle 0|)
I\otimes|-z\rangle_{\bF_2^n}~_{\bF_2^n} \langle -z|)}\nonumber\\
=&
\frac{
\Tr \frac{1}{2^{2(n+l)}}
\sum_{x'',z''\in \bF_2^{n+l}}
(\Lambda^{\pos})^{(x'',z'') }
(|-x_b\rangle \langle -x_b|\otimes 
|0\rangle_{\bF_2^n}~_{\bF_2^n} \langle 0|)
|-x_b\rangle \langle -x_b|\otimes
|-z\rangle_{\bF_2^n}~_{\bF_2^n} \langle -z|)
}{
\Tr \frac{1}{2^{2(n+l)}}
\sum_{x'',z''\in \bF_2^{n+l}}
(\Lambda^{\pos})^{(x'',z'') }
( \rho_{\mix,n}\otimes 
|0\rangle_{\bF_2^n}~_{\bF_2^n} \langle 0|)
I\otimes|-z\rangle_{\bF_2^n}~_{\bF_2^n} \langle -z|)}\nonumber\\
=&
\frac{
\Tr (\Lambda_t)^{\pos}
(|-x\rangle \langle -x|\otimes 
|0\rangle_{\bF_2^n}~_{\bF_2^n} \langle 0|)
|-x_b\rangle \langle -x_b|\otimes
|-z\rangle_{\bF_2^n}~_{\bF_2^n} \langle -z|
}{\Tr
(\Lambda_t)^{\pos}
( \rho_{\mix,n}\otimes 
|0\rangle_{\bF_2^n}~_{\bF_2^n} \langle 0|)
I\otimes|-z\rangle_{\bF_2^n}~_{\bF_2^n} \langle -z|)}\label{1-6-11}\\
=&
\Tr (\Lambda_t)^{\pos,z}
(|-x\rangle \langle -x|)
|-x_b\rangle \langle -x_b|,\nonumber
\end{align}
\end{widetext}
where 
\begin{align*}
&(\Lambda_t)^{\pos,z_c}(\rho) \\
:=&
\sum_{x_k,z_k\in \bF_2^n}
P_{(\Lambda_t)^{\pos},k|Z,c}(x_k, z_k |z_c)
\bX^{x_k} \bZ^{z_k} \rho (\bX^{x_k} \bZ^{z_k})^{\dagger}.
\end{align*}
In the derivation of (\ref{1-6-11}), we use (\ref{1-6-10}).

In this case, we can regard that Bob measures 
the state
$(\Lambda_t)^{\pos,z}(|-x\rangle \langle -x|)$.
Hence, Eve's state can be regarded as
$((\Lambda_t)^{\pos,z})_E(|-x\rangle \langle -x|)$.
Hence, applying Theorem \ref{th-2}, we obtain (\ref{1-6-1}) and (\ref{1-6-2}).

\begin{widetext}
\section{Proof of (\ref{1-6-23}) and (\ref{1-6-23-1})}\label{a1}
First, we evaluate 
$\rE_{\pos}\rE_{z_c}
\rE_{Y_+}
\left[I([z]\in C_1/C_2(Y_+,nh(|z_c|/l+\delta_{|z_c|})
),
\rho_{(\Lambda_t)^{\pos,z},E}^{C_1/C_2(Y)}([z]) )\right]$
as
\begin{align}
& \rE_{\pos}\rE_{z_c}
\rE_{Y_+}
\left[I([z]\in C_1/C_2(Y_+,nh(|z_c|/l+\delta_{|z_c|})
),
\rho_{(\Lambda_t)^{\pos,z},E}^{C_1/C_2(Y)}([z]) )\right]
\nonumber \\
\le &
\rE_{\pos}
\Bigl[
\sum_{|z_c|< \underline{k}}
P_{(\Lambda_t)^{\pos},Z,c}(z_c)
\overline{h}
\Bigl(
\sum_{k_k=0}^n
\tilde{P}_{(\Lambda_t)^{\pos},Z}(k_k |z_c)
f(k_k,\underline{k}|n,l,\delta_{\underline{k}})
\Bigr)
\nonumber \\
&\quad + 
\sum_{\underline{k}\le |z_c|\le \overline{k}}
P_{(\Lambda_t)^{\pos},Z,c}(z_c)
\overline{h}
\Bigl(
\sum_{k_k=0}^n
\tilde{P}_{(\Lambda_t)^{\pos},Z}(k_k |z_c)
f(k_k,|z_c|~|n,l,\delta_{|z_c|})
\Bigr)\Bigr]\nonumber\\
&+
\rE_{\pos}
\Bigl[
\sum_{|z_c|< \underline{k}}
P_{(\Lambda_t)^{\pos},Z,c}(z_c)
n(R-h(\underline{k}/l+\delta_{\underline{k}}))
\sum_{k_k=0}^n
\tilde{P}_{(\Lambda_t)^{\pos},Z}(k_k |z_c)
f(k_k,\underline{k}|n,l,\delta_k)\nonumber \\
&\quad +\sum_{\underline{k}\le |z_c|\le \overline{k}}
P_{(\Lambda_t)^{\pos},Z,c}(z_c)
n(R-h(|z_c|/l+\delta_{|z_c|}))
\sum_{k_k=0}^n
\tilde{P}_{(\Lambda_t)^{\pos},Z}(k_k |z_c)
f(k_k,|z_c|~|n,l,\delta_{|z_c|})
\Bigr]\label{1-6-4}\\
\le &
\overline{h}
\Bigl(
\rE_{\pos}
\Bigl[
\sum_{|z_c|< \underline{k}}
P_{(\Lambda_t)^{\pos},Z,c}(z_c)
\sum_{k_k=0}^n
\tilde{P}_{(\Lambda_t)^{\pos},Z}(k_k |z_c)
f(k_k,\underline{k}|n,l,\delta_{\underline{k}})
\nonumber \\
&\quad + 
\sum_{\underline{k}\le |z_c|\le \overline{k}}
P_{(\Lambda_t)^{\pos},Z,c}(z_c)
\sum_{k_k=0}^n
\tilde{P}_{(\Lambda_t)^{\pos},Z}(k_k |z_c)
f(k_k,|z_c|~|n,l,\delta_{|z_c|})
\Bigr]
\Bigr)\nonumber\\
&+
\rE_{\pos}
\Bigl[
\sum_{|z_c|< \underline{k}}
P_{(\Lambda_t)^{\pos},Z,c}(z_c)
n(R-h(\underline{k}/l+\delta_{\underline{k}}))
\sum_{k_k=0}^n
\tilde{P}_{(\Lambda_t)^{\pos},Z}(k_k |z_c)
f(k_k,\underline{k}|n,l,\delta_k)\nonumber \\
&\quad +\sum_{\underline{k}\le |z_c|\le \overline{k}}
P_{(\Lambda_t)^{\pos},Z,c}(z_c)
n(R-h(|z_c|/l+\delta_{|z_c|}))
\sum_{k_k=0}^n
\tilde{P}_{(\Lambda_t)^{\pos},Z}(k_k |z_c)
f(k_k,|z_c|~|n,l,\delta_k)
\Bigr]\label{1-6-5}\\
= &
\overline{h}
\Bigl(
\sum_{k_k=0}^n
\sum_{k_c=0}^{\underline{k}}
\rE_{\pos}
\Bigl[
\tilde{P}_{(\Lambda_t)^{\pos},Z,k,c}(k_k,k_c)
f(k_k,\underline{k}|n,l,\delta_{\underline{k}})\Bigr]
\nonumber \\
&\quad + 
\sum_{k_k=0}^n
\sum_{k_c=\underline{k}+1}^{\overline{k}}
\rE_{\pos}
\Bigl[
\tilde{P}_{(\Lambda_t)^{\pos},Z,k,c}(k_k,k_c)
f(k_k,k_c|n,l,\delta_{k_c})
\Bigr]\Bigr)\nonumber\\
&+
\sum_{k_k=0}^n
\sum_{k_c=0}^{\underline{k}}
\rE_{\pos}
\Bigl[
\tilde{P}_{(\Lambda_t)^{\pos},Z,k,c}(k_k,k_c)
n(R-h(\underline{k}/l+\delta_{\underline{k}}))
f(k_k,\underline{k}|n,l,\delta_{\underline{k}})
\Bigr]
\nonumber \\
&\quad + 
\sum_{k_k=0}^n
\sum_{k_c=\underline{k}+1}^{\overline{k}}
\rE_{\pos}
\Bigl[
\tilde{P}_{(\Lambda_t)^{\pos},Z,k,c}(k_k,k_c)
n(R-h(k_c/l+\delta_{k_c}))
f(k_k,k_c|n,l,\delta_{k+c})
\Bigr]
\label{2-15-1}
\end{align}
Further, RHS of (\ref{2-15-1}) is evaluated as
\begin{align}
&\hbox{(RHS of (\ref{2-15-1}))}\nonumber \\
=&
\overline{h}
\Bigl(
\sum_{k_k=0}^n
\sum_{k_c=0}^{\underline{k}}
\tilde{P}_{(\Lambda_t),Z}(k_k+k_c)P_{hg}(k_c|n,l,k_k+ k_c)
f(k_k,\underline{k}|n,l,\delta_{\underline{k}})
\nonumber \\
&\quad + 
\sum_{k_k=0}^n
\sum_{k_c=\underline{k}+1}^{\overline{k}}
\tilde{P}_{(\Lambda_t),Z}(k_k+k_c)P_{hg}(k_c|n,l,k_k+ k_c)
f(k_k,k_c|n,l,\delta_{k_c})
\Bigr)\nonumber\\
&+
\sum_{k_k=0}^n
\sum_{k_c=0}^{\underline{k}}
\tilde{P}_{(\Lambda_t),Z}(k_k+k_c)P_{hg}(k_c|n,l,k_k+ k_c)
n(R-h(\underline{k}/l+\delta_{\underline{k}}))
f(k_k,\underline{k}|n,l,\delta_{\underline{k}})
\nonumber \\
&\quad + 
\sum_{k_k=0}^n
\sum_{k_c=\underline{k}+1}^{\overline{k}}
\tilde{P}_{(\Lambda_t),Z}(k_k+k_c)P_{hg}(k_c|n,l,k_k+ k_c)
n(R-h(k_c/l+\delta_{k_c}))
f(k_k,k_c|n,l,\delta_{k+c})\label{1-6-6}\\
\le &
\overline{h}
\Bigl(
\max_j 
\Bigl[
\sum_{k_c=0}^{\underline{k}}
P_{hg}(k_c|n,l,j)
f(k_k,\underline{k}|n,l,\delta_{\underline{k}})
+
\sum_{k_c=\underline{k}+1}^{\overline{k}}
P_{hg}(k_c|n,l,j)
f(k_k,k_c|n,l,\delta_{k_c})
\Bigr]\Bigr)\nonumber\\
&+
\max_j 
\Bigl[
\sum_{k_c=0}^{\underline{k}}
P_{hg}(k_c|n,l,j)
n(R-h(\underline{k}/l+\delta_{\underline{k}}))
f(k_k,\underline{k}|n,l,\delta_{\underline{k}})
\nonumber \\
&\hspace{10ex} + 
\sum_{k_c=\underline{k}+1}^{\overline{k}}
P_{hg}(k_c|n,l,j)
n(R-h(k_c/l+\delta_{k_c}))
f(k_k,k_c|n,l,\delta_{k+c})
\Bigr].\label{1-6-3}
\end{align}

In the above relations,
(\ref{1-6-4}) follows from (\ref{1-6-1}) and (\ref{1-6-2}),
and 
(\ref{1-6-5}) follows from the convexity of $\bar{h}$,
(\ref{1-6-6}) follows from (\ref{1-6-7}),
(\ref{1-6-3}) follows by replacing $k_k+ k_c$ by $j$.
Hence, we obtain (\ref{1-6-23}).

Similarly, we have
\begin{align}
& \rE_{\pos}\rE_{z_c}
\rE_{Y_+}
\Bigl[
\frac{I([z]\in C_1/C_2(Y_+,nh(|z_c|/l+\delta_{|z_c|})
),\rho_{(\Lambda_t)^{\pos,z},E}^{C_1/C_2(Y)}([z]) )}
{n(R-h(k_\times/l_\times +\delta_{k_\times}))}\Bigr]
\nonumber \\
\le &
\frac{1}{n(R-h(\overline{k}/l_\times +\delta_{\overline{k}}))}
\rE_{\pos}
\Bigl[
\sum_{|z_c|< \underline{k}}
P_{(\Lambda_t)^{\pos},Z,c}(z_c)
\overline{h}
\Bigl(
\sum_{k_k=0}^n
\tilde{P}_{(\Lambda_t)^{\pos},Z}(k_k |z_c)
f(k_k,\underline{k}|n,l,\delta_{\underline{k}})
\Bigr)\nonumber \\
&\quad + 
\sum_{\underline{k}\le |z_c|\le \overline{k}}
P_{(\Lambda_t)^{\pos},Z,c}(z_c)
\overline{h}
\Bigl(
\sum_{k_k=0}^n
\tilde{P}_{(\Lambda_t)^{\pos},Z}(k_k |z_c)
f(k_k,|z_c|~|n,l,\delta_{|z_c|})
\Bigr)
\Bigr]\nonumber\\
&+
\rE_{\pos}
\Bigl[
\sum_{|z_c|< \underline{k}}
P_{(\Lambda_t)^{\pos},Z,c}(z_c)
\sum_{k_k=0}^n
\tilde{P}_{(\Lambda_t)^{\pos},Z}(k_k |z_c)
f(k_k,\underline{k}|n,l,\delta_k)\nonumber \\
&\quad +\sum_{\underline{k}\le |z_c|\le \overline{k}}
P_{(\Lambda_t)^{\pos},Z,c}(z_c)
\sum_{k_k=0}^n
\tilde{P}_{(\Lambda_t)^{\pos},Z}(k_k |z_c)
f(k_k,|z_c|~|n,l,\delta_{|z_c|})
\Bigr]\nonumber\\
\le &
\frac{1}{n(R-h(\overline{k}/l_\times +\delta_{\overline{k}}))}
\overline{h}
\Bigl(
\max_j 
\Bigl[
\sum_{k_c=0}^{\underline{k}}
P_{hg}(k_c|n,l,j)
f(k_k,\underline{k}|n,l,\delta_{\underline{k}})
+
\sum_{k_c=\underline{k}+1}^{\overline{k}}
P_{hg}(k_c|n,l,j)
f(k_k,k_c|n,l,\delta_{k_c})
\Bigr]\Bigr)\nonumber\\
&+
\max_j 
\Bigl[
\sum_{k_c=0}^{\underline{k}}
P_{hg}(k_c|n,l,j)
f(k_k,\underline{k}|n,l,\delta_{\underline{k}})
+\sum_{k_c=\underline{k}+1}^{\overline{k}}
P_{hg}(k_c|n,l,j)
f(k_k,k_c|n,l,\delta_{k+c})
\Bigr].\nonumber
\end{align}
Hence, we obtain (\ref{1-6-23-1}).
\end{widetext}


\begin{thebibliography}{99}
\bibitem{bene} C.H. Bennett and G. Brassard, 
``Quantum cryptography: Public key distribution and coin tossing,''
Proc. IEEE Int. Conf. on Computers, Systems, and Signal Processing 
(Bangalore, India, IEEE, New York, 1984) 175.

\bibitem{mayer1} D. Mayers, 
``Quantum key distribution and string oblivious transfer in noisy channels,''
In {\em Advances in Cryptology -- Proc. Crypto'96, Vol. 1109 of Lecture Notes in Computer Science} 
(Ed. N. Koblitz, Springer-Verlag, New York, 1996) 343; 
J. Assoc. Comput. Mach. 48 (2001) 351.

\bibitem{ILM}
H. Inamori, N. L\"{u}tkenhaus, and D. Mayers
``Unconditional Security of Practical Quantum Key Distribution,''
quant-ph/0107017.

\bibitem{shor2} P. W. Shor and J. Preskill, 
``Simple Proof of Security of the BB84 Quantum Key Distribution Protocol,''
{\em Phys. Rev. Lett.} {\bf 85}, 441 (2000).

\bibitem{C-S}
A. R. Calderbank and P. W. Shor, 
``Good quantum error correcting codes exist,'' 
{\em Phys. Rev. A}, {\bf 54}, 1098 -- 1105 (1996).

\bibitem{Steane}
M. Steane, 
``Multiple particle interference and quantum error correction,''
{\em Proc. Roy. Soc. Lond. A}, {\bf 452}, 2551 -- 2577 (1996).

\bibitem {Hamada03} M. Hamada, 
``Reliability of Calderbank-Shor-Steane Codes and Security of Quantum Key Distribution,''
J. Phys. A: Math. Gen. {\bf 37}, (2004).

\bibitem{CRE}
M. Christandl, R. Renner, and A. Ekert,
``A Generic Security Proof for Quantum Key Distribution'',
quant-ph/0402131v2.

\bibitem{RGK}
R. Renner, N. Gisin and B. Kraus,
``Information-theoretic security proof for quantum-key-distribution
protocols,''
{\it Phys. Rev}. {\bf A72} (2005) 012332,
quant-ph/0502064.

\bibitem{Koashi}
M. Koashi,
``Simple security proof of quantum key distribution via uncertainty 
principle,'' quant-ph/0505108.

\bibitem{WMU}
S. Watanabe, R. Matsumoto, T. Uyematsu,
``Noise Tolerance of the BB84 Protocol with Random Privacy Amplification,''
quant-ph/0412070.

\bibitem{SGGRZ}
D. Stucki, N. Gisin, O. Guinnard, G. Ribordy, and H. Zbinden,
``Quantum key distribution over 67 km with a plug \& play system,''
{\em New J. Phys.}, {\bf 4}, 41 (2002).

\bibitem{KNHTKN04}
T. Kimura, Y. Nambu, T. Hatanaka, A. Tomita, H.
Kosaka, and K. Nakamura, 
{\em Jpn. J. Appl. Phys.}, {\bf 43}, L1217 (2004).

\bibitem{GYS}
C. Gobby, Z. L. Yuan and A. J. Shields,
{\em Appl. Phys. Lett.}, {\bf 84}, 3762 (2004).

\bibitem{TMTT05} 
A. Tanaka, W. Maeda, A. Tajima, and S. Takahashi,
\textit{Proceedings of the 18th Annual Meeting of the IEEE Lasers and Electro-Optics Society}, Sidney, Australia, 23--27 October 2005, p.\,557

\bibitem{Y-S}
Z. L. Yuan and A. J. Shields,
``Continuous operation of a one-way quantum key
distribution system over installed telecom fiber,''
{\em Optics Express}, {\bf 13}, 660 (2005).

\bibitem{K-P}
M. Koashi and J. Preskill, 
{\em Phys. Rev. Lett.}, {\bf 90}, 057902 (2003).

\bibitem{YMI}
Y. Watanabe, W. Matsumoto, and Hideki Imai,
``Information reconciliation in quantum key distribution
using low-density parity-check codes,''
{\em Proc. of 
International Symposium on Information Theory and its Applications, 
ISITA2004, Parma, Italy, October, 2004}, p. 1265 -- 1269.

\bibitem{C-K}
Csisz\'{a}r and J. K\"{o}rner, 
{\em Information Theory: Coding Theorems for Discrete Memoryless Systems}. 
(NY: Academic, 1981).

\bibitem{weyl}
H. Weyl, 
{\em Gruppentheorie und Quantenmechanik}, 
(Leipzig: Verlag von S. Hirzel, 1928). 
English translation, 
{\em The Theory of Groups and Quantum Mechanics}, 
of the second (1931) ed. was reprinted by Dover, 1950.

\bibitem{BDSW}
C. H. Bennett, D. P. DiVincenzo, J. A. Smolin, and W. K. Wootters, 
``Mixed-state entanglement and quantum error correction,'' 
{\em Phys. Rev. A}, {\bf 54}, 3824 -- 3851 (1996).

\bibitem{Hama2}
M. Hamada, 
``Teleportation and entanglement distillation in the presence of correlation among bipartite mixed states,'' 
{\em Phys. Rev. A}, {\bf 68}, 012301 (2003).

\bibitem{hwang} W-Y. Hwang, 
``Quantum Key Distribution with High Loss: Toward Global Secure Communication,''
{\em Phys. Rev. Lett.}, {\bf 91}, 057901 (2003).

\bibitem{GLLP}
D. Gottesman, H.-K. Lo, N. L\"{u}tkenhaus, and J. Preskill, 
``Security of quantum key distribution with imperfect devices,''
{\em Quant. Inf. Comput.}, {\bf 5}, 325 -- 360 (2004). 

\bibitem{lo4} H.-K. Lo, 
``Quantum Key Distribution with Vacua or Dim Pulses as Decoy States,''
{\em Proc. 2004 IEEE Int. Symp. on Inf. Theor.} 
(June 27-July 2, Chicago, 2004) 17.

\bibitem{lolo} H. K. Lo, X.-F. Ma, and K. Chen, 
``Decoy State Quantum Key Distribution,''
{\em Phys. Rev. Lett.}, {\bf 94}, 230504, (2005).

\bibitem{wang2d} X.-B. Wang, 
``Beating the PNS attack in practical quantum cryptography,''
{\em Phys. Rev. Lett.}, {\bf 94}, 230503 (2005).

\end{thebibliography}
\end{document}